\documentclass[referee]{raa}           

\usepackage{graphicx,times}
\usepackage{natbib}
\usepackage{amssymb,amsmath}
\bibpunct{(}{)}{;}{a}{}{,}

\usepackage[pagebackref=true]{hyperref}

\begin{document}

   \title{Interstellar extinction, polarization efficiency, and grain alignment in the direction towards bright-rimmed clouds and cometary globules}

 \volnopage{ {\bf 20XX} Vol.\ {\bf X} No. {\bf XX}, 000--000}
   \setcounter{page}{1}

   \author{Saikhom Pravash
   \inst{1,2}
   }

   \institute{Indian Institute of Astrophysics, II Block, Koramangala, 560034, India; {\it saikhom.singh@iiap.res.in}\\
        \and
             Pondicherry University, R.V. Nagar, Kalapet, Puducherry, 605014, India\\
\vs \no
   %{\small Received 20XX Month Day; accepted 20XX Month Day}
}

\abstract{The polarization of starlight and thermal dust emission, resulting from non-spherical grains aligned with the interstellar magnetic field (B-field), act as a powerful tool to trace the B-field morphologies and strengths in molecular clouds and constrain the grain alignment mechanisms and grain properties. The exact alignment mechanisms of grains is not yet fully clear. However, the leading theory is the alignment induced by RAdiative Torques (RATs), known as RAT theory. In this work, we use optical polarization observations of background stars projected towards nine of Bright-Rimmed Clouds (BRCs) and Cometary Globules(CGs) to study the polarization efficiencies and the alignment mechanisms of the grains in the direction towards the outer diffuse envelopes of these clouds. We use distance and extinction data of the stars from Gaia EDR3 and StarHorse 2 Catalogue. We study the variations of the degree and position angle of polarization, and the extinction, as functions of distance of the stars. For some of the clouds, we find discrete enhancement of the extinction at certain distances along with an increase in polarization degree, signifying the presence of polarizing dust layers. We estimate the polarization efficiency of grains towards each of the clouds. We find that it decreases with increasing extinction, and also shows a slight increase with dust temperature for some clouds associated with more ordered magnetic field orientations, providing an implication for the alignment of grains by RATs. Whereas, for some other clouds, the decrease in the polarization efficiency with extinction may be caused by more fluctuations in the magnetic field orientations.
\keywords{ISM: clouds -- ISM: dust, extinction -- ISM: magnetic fields
}
}

   \authorrunning{Pravash}           
   \titlerunning{Polarization efficiency towards BRCs and cometary globules}
   \maketitle

%\linenumbers

%%%%%%%%%%%%%%%%%%%%%%%%%%%%%%%%%%%%%%%%%%%%%%%%%%

%%%%%%%%%%%%%%%%% BODY OF PAPER %%%%%%%%%%%%%%%%%%

\section{Introduction}\label{section:Introduction}

The dust in the interstellar medium is crucial element that is involved in various processes including star and planet formation, gas heating and cooling, in providing surface for the formation of different complex molecules (see \citealt{2003ARA&A..41..241D}), and in the tracing of magnetic fields. 

When starlight passes through the dusty interstellar medium it becomes partially plane polarized up to a few percent, as first observed by \citet{1949ApJ...109..471H} and \cite{1949Sci...109..166H}.  This requires that three conditions are met (i) the individual dust particles must be optically anisotropic (ii) there must be net alignment of the grain axes of anisotropy with one of their principal axes ("internal alignment"), and (iii) the spin axes of the individual grains must be collectively aligned with an external reference direction("external alignment"). The grain anisotropy most likely comes from asymmetry in the shape of the dust particles. The observed polarization is then due to dichroic extinction by the asymmetric dust grains aligned with the local interstellar magnetic field \citep{1949Sci...109..165H, 1951ApJ...114..206D}. The dust grains are aligned in such a way that their short axes are parallel to the magnetic field. The grains also emit polarized thermal radiation at longer wavelengths. For starlight polarization, the polarization angle traces the plane-of-sky projection of the  magnetic field.  Different possible mechanisms for the physics of grain alignment have been suggested \citep[e.g,][]{1951ApJ...114..206D, 1952MNRAS.112..215G, 2007MNRAS.378..910L}.

The currently most accepted mechanism of grain alignment is based on the RAdiative Torques (RATs) theory, first introduced by \cite{1976Ap&SS..43..291D} and numerically demonstrated by \cite{1997ApJ...480..633D}. An analytical RAT theory was later developed by \cite{2007MNRAS.378..910L} and \cite{2008MNRAS.388..117H}. In RAT theory, an irregular dust grain is spun up when anisotropic radiation with $\lambda < 2a$ (where $\lambda$ is the wavelength of the incident radiation and $a$ is the average grain size) is incident on the grain.  This results from the radiative torques due to the difference in the scattering cross-sections for the left and right circular polarization components of the incident radiation (\citealt{1997ApJ...480..633D}; \citealt{2007MNRAS.378..910L}). These radiative torques can be very efficient and allow the grains to rotate with rotation rates significantly exceeding thermal rates (suprathermal rotation) (\citep{2008MNRAS.388..117H, 2015ARA&A..53..501A, 2016ApJ...831..159H}. In addition, grain alignment with the magnetic field requires the grains to be, at least, paramagnetic.  In the ISM this mainly corresponds to silicate grains.  Dust-induced polarization can, therefore, trace the grain alignment efficiency and the properties of the grains, and trace the orientation of the plane-of-sky projection of the magnetic field \citep[][and ref.s therein]{2015ARA&A..53..501A}.  

In this work, we use archival R-band polarization observations of background stars projected towards the outer envelopes of several Bright-Rimmed Clouds (BRCs) and Cometary Globules (CGs), probing the diffuse regions towards the envelopes of the clouds. These types of clouds are irradiated by ionizing sources (O and B type stars mainly) located at varying distances from the cloud surfaces, causing enhanced grain illumination and photodissociation of the gas. 
Employing such well-characterized source geometries allow us to study the role of radiation field strength and color on the grain alignment. Previous studies of these effects include \cite{2019ApJ...873...87M}, who
analyzed the polarization in the Local Bubble wall, and \cite{2021AJ....161..149S} who probed the reflection nebulae in Sh2-185. Those studies investigated the quantitative relationship between the observed polarization efficiency and the radiation field strength, as well as the strength of magnetic fields. In this study, we will focus mainly on the polarization efficiency of the grains in the direction towards the diffuse outer regions of the clouds and its variation as a function of extinction and dust temperature to study the grain alignment mechanisms. We do not directly address the relationship between polarization efficiency and the luminosity of the ionizing stars.

BRCs and CGs are different from diffuse and dark clouds as they are illuminated by strong anisotropic radiation fields from nearby O and B type stars. For diffuse and dark clouds, the radiation field is dominated by diffuse interstellar radiation field if there are no embedded sources inside the cloud. 

We chose the bright-rimmed clouds BRCs IC1396A, BRC37, BRC38 and BRC39 for our study. These are all located in the boundary of the IC 1396 H II region with the common ionization source HD 206267, of spectral class O6.5 V.  In addition, their magnetic field morphologies have been mapped \citep{2018MNRAS.476.4782S}. Therefore, these BRCs constitute a good target group for studying the variations in polarization efficiency in a cloud complex with a common ionization source.

Additionally, we also choose some cometary globules LDN323 (hereafter L323), LDN328 (hereafter L328), LDN331 (hereafter L331), LBN437 and GAL110-13 for the study apart from the chosen BRCs. Cometary globules are evolved stages of BRCs and found in isolation. However, BRCs can be found on the peripheries of HII regions. The cometary globules L323, L328 and L331 form same complex and three closest B-type stars are considered to be the common ionizing sources (see section 1.2). LBN437 and GAL110-13 are separate clouds but their main ionizing source is the same i.e 10 Lac (see section 1.2). The magnetic field morphologies for the outer envelopes of L323, L328 and L331 have been studied by \citealt{2023MNRAS.524.1219K}, for LBN437 by \citealt{2013MNRAS.432.1502S} and for GAL110-13 by \citealt{2016A&A...588A..45N}.

Further details about the clouds are given below in sections \ref{subsection:BRCs} and \ref{subsection:CGs}. The paper is organized as follows: section \ref{section:Data acquisition} shows data acquisition details, section \ref{section:Analysis} presents the data analysis and results, section \ref{section:Discussions} presents discussion of the results and section \ref{section:Summary} gives summary of the work.

\subsection{\it{Bright Rimmed Clouds IC1396A, BRC37, BRC38, BRC39}\label{subsection:BRCs}}
IC1396A, BRC37, BRC38 and BRC39 are four BRCs which show different structures found at the boundary of the HII region IC1396 located in the Cep OB2 association \citep{1991ApJ...370..263S}. The expansion of this HII region has swept up a molecular ring of a radius of nearly 12 pc \citep{1995ApJ...447..721P}. IC1396 is estimated to be at a distance of 750pc \citep{1976PASP...88..865G}. The cloud BRC38 (a.k.a. IC1396N) is located in the North, BRC37 in the South, BRC39 in the East and IC1396A is in the West of the IC1396 HII region. The dominant source of UV radiation, which acts as a primary ionizing source of the clouds, is the star HD206267 with spectral type O6.5V, located nearly at the center of the HII region (\citealt{1984ApJ...286..718W} ; \citealt{1995ApJ...447..721P}). The clouds IC1396A, BRC37, BRC38, and BRC39 are at projected distances of nearly 4.9 pc, 12.0 pc, 10.6 pc, and 12.6 pc, respectively, from HD206267 assuming a distance to IC1396 of 750 pc.

\subsection{\it{Cometary Globules L323, L328, L331, LBN437, GAL110-13}\label{subsection:CGs}}

The cometary globules L323, L328 and L331 are part of same complex but with different cloud morphologies.  The clouds are located at a distance of about 220pc \citep{2011A&A...536A..99M}.  The three closely grouped B-type stars HD167863, HD168675 and HD168352, lie at angular distances of 47.5 arcmin, 1.11$^\circ$ and 1.18$^\circ$, respectively, from the three clouds \citep{2023MNRAS.524.1219K}. 

L328 has a dark opaque region of nearly 3$\times$3 arcmin$^2$ and many long curved structures with sizes of up to nearly 11 arcmin. In the morphology of this cloud, the tail extends in projection on the sky to nearly 15 arcmin. L331, which is located around 10 arcmin to the north of L328, has a size of 3$\times$3   arcmin$^2$ and an area of 0.005 deg$^2$ \citep{1962ApJS....7....1L}. L323, which is located 20 arcmin to the west of L328, has a size of 11$\times$11  arcmin$^2$ and an area of 0.016 deg$^2$ \citep{1962ApJS....7....1L}.

The cometary globule LBN437 (also known as Gal 96-15) \citep{1988BAAS...20..957O} is located on the border of the HII region Sh2-126 \citep{1959ApJS....4..257S}. This cometary globule is located at a distance of 360 $\pm$ 65 pc \citep{2013MNRAS.432.1502S}. The Lac OB1 association, dominated by the O9 V star 10 Lac is considered responsible for the cometary shape of LBN437 \citep{1994A&A...290..235O}. 

GAL110-13, also known as LBN534 or DG191, is an isolated, unusually elongated, comet-shaped molecular cloud located at a distance of 450 $\pm$ 80pc \citep{2016A&A...588A..45N}. The major axis of the cloud is $\approx 1^\circ$ in extent and it has an orientation with a position angle of $\approx 45^\circ$ to the east with respect to the north. The cloud has a width of ~8 arcminute. 10 Lac, a member of the Lac OB1 association, is the main ionizing star which might have compressed the cloud and creates the cometary morphology and subsequent star formation  \citep{2007ApJ...657..884L}. Taking a distance of 450 pc to GAL110-13, the spatial separation between the cloud and 10 Lac is $\approx$ 110 pc. 

\section{Data acquisition}\label{section:Data acquisition}
\subsection{\it{Archival polarization data}}\label{subsection:Polarimetry}
In this work, we have used archival optical polarization data towards these BRCs and CGs published in many recent studies. The polarization data for IC1396A, BRC37, BRC38 and BRC39 are from \cite{2018MNRAS.476.4782S}, those for L323, L328 and L331 from \cite{2023MNRAS.524.1219K}, those for LBN437 from \cite{2013MNRAS.432.1502S} and those for GAL110-13 from \cite{2016A&A...588A..45N}. For each of these clouds, the observations were acquired at the 104 cm Sampurnanand Telescope of the Aryabhatta Research Institute of Observational Sciences (ARIES). The polarimetry was acquired in R-band, using the ARIES IMaging POLarimeter(AIMPOL) (\citealt{2004BASI...32..159R}) mounted at the Cassegrain focus of the telescope. The instrument uses a TK 1024 $\times$ 1024 pixel$^2$ CCD Camera and an achromatic half-wave plate modulator together with a Wallaston Prism Beam splitter. The CCD yields a plate scale of 1.48 arcsec pixel$^{-1}$ and a field of view of around 8 arcminutes. The image Full Width at Half-Maximum (FWHM) varies from 2 to 3 pixels and the read-out noise and the gain of CCD are 7 e$^{-1}$ and 11.98 e$^{-1}$ ADU$^{-1}$ respectively. The details of observations and data reduction for each region can be found in the respective references cited above. Because of the nature of polarization uncertainties \citep[e.g.][]{2006PASP..118.1340V} we use only polarization data satisfying signal-to-noise ratio of $P/\sigma_P > 2$ in our analysis.  %%We converted these R-band polarization (P) data into V-band, $\mathrm{P_V}$ using the relations of $\lambda_{max}$ and $A_V$ given in \citet{2007ApJ...665..369A} and Serkowski law.

\subsection{\it{Ancillary data }}\label{subsection:Ancillary data}
We use the StarHorse2 catalogue based on Gaia EDR3 \citep[for more details see][]{2022A&A...658A..91A} for the distances and extinction values of the background stars, projected towards the different clouds of our study having polarization measurements. StarHorse \citep{2018MNRAS.476.2556Q} is an isochrone-fitting code and a Bayesian tool used to determine distances, extinctions (at $\lambda$ = 542nm) $A_V$, ages, masses, effective temperatures, metallicities and surface gravities for field stars.

\section{Analysis and Results}\label{section:Analysis}
For the distance data of the stars as given in the catalogue based on Gaia EDR3, we use $dist50$ which means 50th percentile distance as the distance (hereafter denoted by $d$ in further analysis) of the star since it is the median value and $dist16$ (16th percentile distance) and dist84 (84th percentile distance) for estimating the uncertainties in distance of the stars. We take ($dist84 - dist16$)/2 as the 1 $\sigma$ uncertainty in the distance for each of the stars. Similarly, for the extinction $A_V$ values we use $A_V50$ which means the line-of-sight extinction at $\lambda=5420($\AA$)$ 50th percentile as the extinction (hereafter denoted by $A_V$ in further analysis) values of each of the stars since it is the median value. $A_V16$ (16th percentile) and $A_V84$ (84th percentile) are used to estimate 1 $\sigma$ uncertainties in $A_V$ values of the stars given by ($A_V84-A_V16$)/2.

The $A_V$ values in our sample range from around 1 to 5 mag, which corresponds to diffuse and translucent regimes as well as the outer envelopes of molecular clouds. In these environments, dust grains are not expected to undergo significant growth, and the grain properties and the grain size distribution likely remain similar to that of the diffuse interstellar medium. The total-to-selective extinction ratio $R_V$ provides the information on the average size of grains along the line of sight, with higher values implying larger grain sizes, and the typical value for the diffuse interstellar medium is 3.1. Since we probe the outer diffuse regions of the clouds, which are likely similar to the diffuse interstellar medium, we take $R_V = 3.1$ to estimate the $\lambda_{max}$ value using the relation given in \cite{2007ApJ...665..369A}. Then we apply the Serkowski relation \citep{1975ApJ...196..261S} to convert the observed R-band polarization data $P_R$ to V-band $P_V$ for a consistent analysis and comparison with $A_V$ in the same band. In the conversion using the Serkowski relation, we take the values of the estimated $\lambda_{max}$, the $\lambda$ values of the corresponding bands, and the width of the polarization curve parameter K = 1.15, which is an empirically derived value and valid for diffuse ISM conditions. We also note that the $R_V$ values could be subject to change with $A_V$, with its value expected to be slightly increased for the regions of higher $A_V$ ($4-5$ mag). However, the resulting uncertainties from the assumption of $R_V = 3.1$ in the estimation of $\lambda_{max}$ values are expected to be small, given the range of $A_V$ in our study, and would not significantly affect our results.

\begin{figure*}
    \centering
        \includegraphics[scale=0.3]{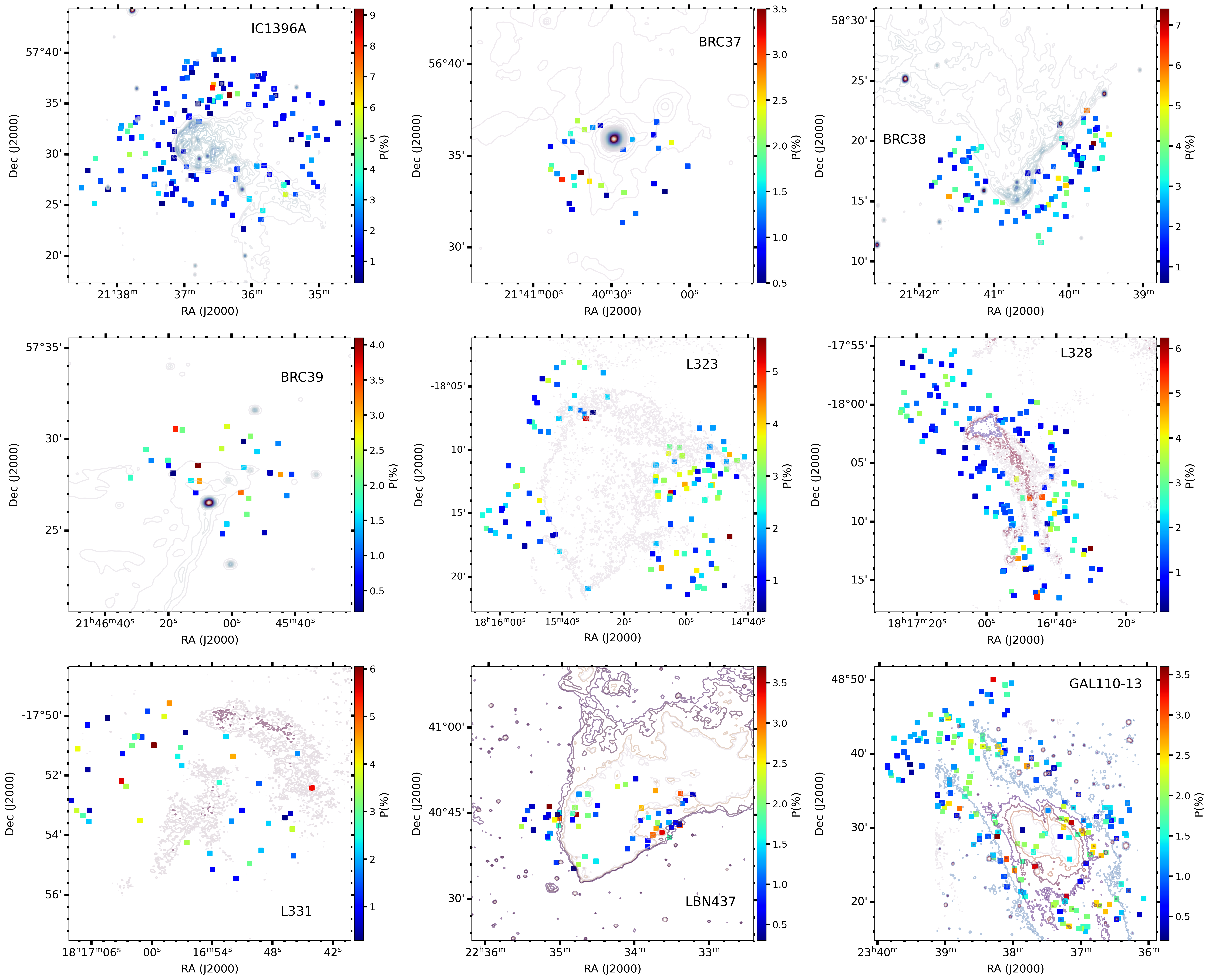}
    \caption{Spatial distribution maps of degree of polarization $P$ for the stars projected in the direction towards the clouds IC1396A, BRC37, BRC38, BRC39, L323, L328, L331, LBN437 and GAL110-13. The gray contours for IC1396A, BRC37, BRC38, BRC39, LBN437 and GAL110-13 are the WISE 12$\mu$m contours of the respective clouds and the gray contours for L323, L328 and L331 are the Digitised Sky Survey(DSS)-2-Red contours of the respective clouds.}
    \label{Fig:P_map}
\end{figure*}

\begin{figure*}
    \centering
        \includegraphics[scale=0.3]{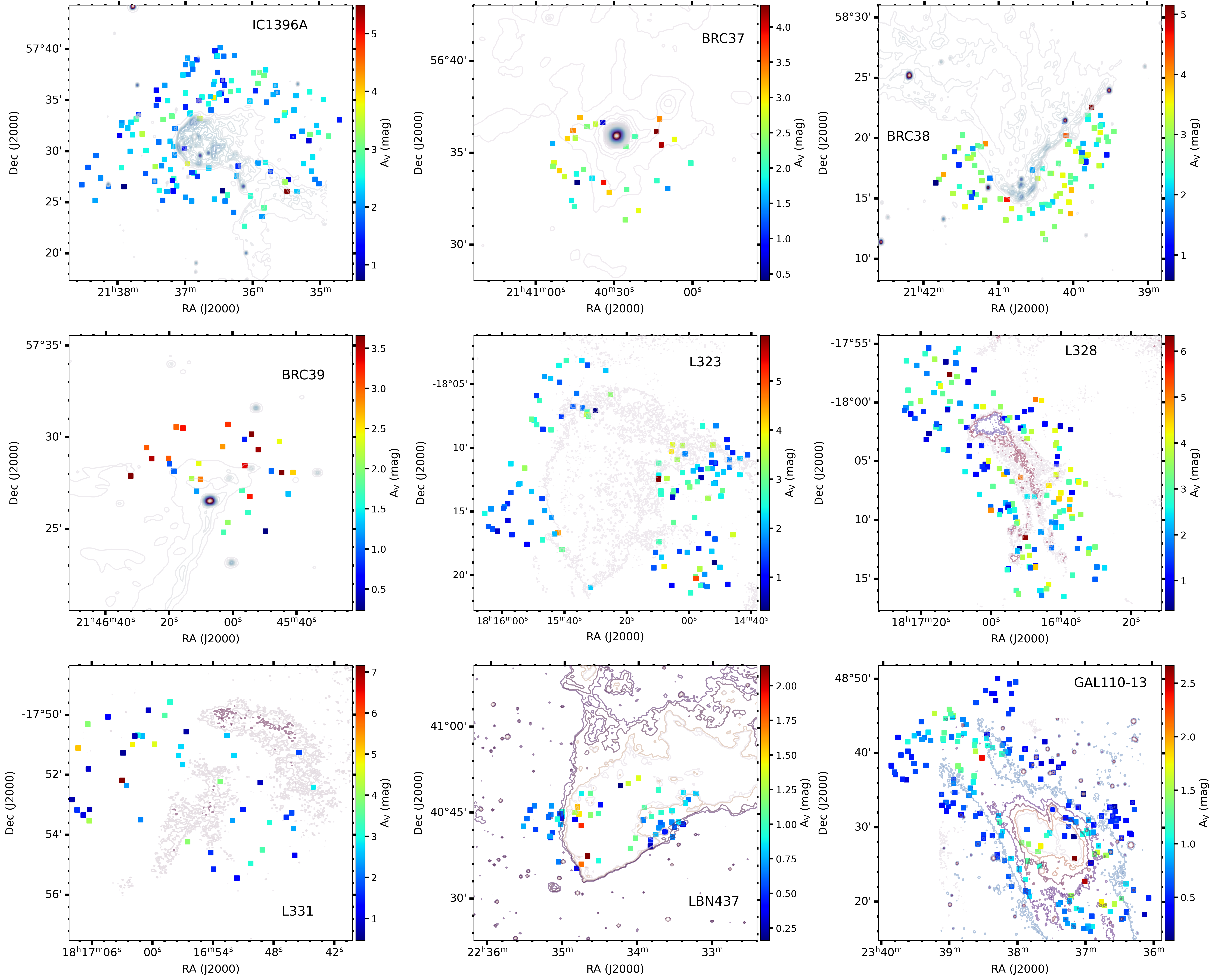}
    \caption{Spatial distribution maps of extinction $A_V$ observed towards each of the clouds. The gray contours are the same as in Figure \ref{Fig:P_map}.} 
    \label{Fig:Av_map}
\end{figure*}

\begin{figure*}
    \centering
        \includegraphics[scale=0.3]{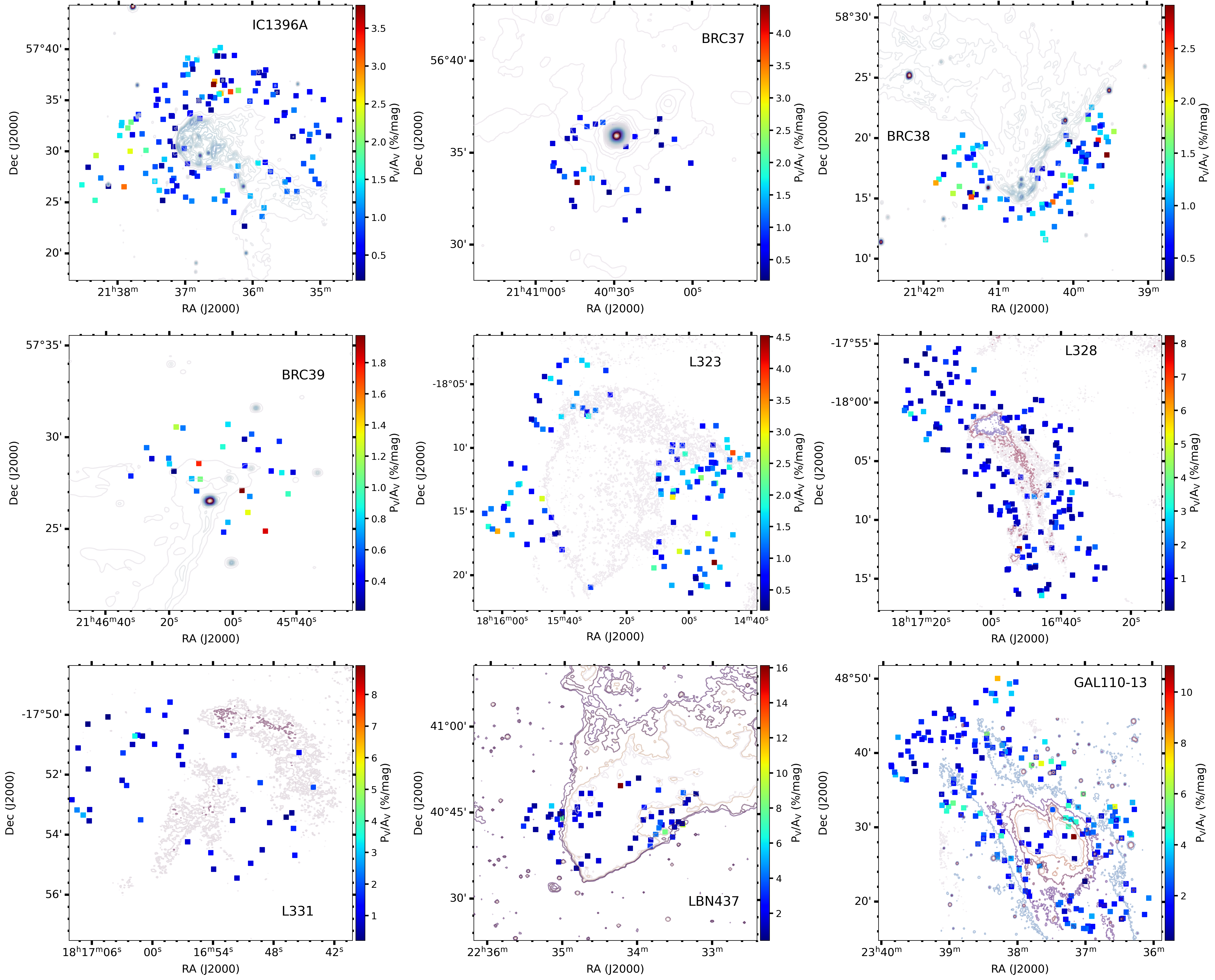}
    \caption{Spatial distribution maps of polarization efficiency $\frac{P_V}{A_V}$ measured towards each of the clouds. The gray contours are the same as in Figure \ref{Fig:P_map}.} 
    \label{Fig:Pv_Av_map}
\end{figure*}

\subsection{\it{Maps of polarization degree, amount of extinction and polarization efficiency}}
Figure \ref{Fig:P_map} shows the maps of spatial distribution of the degree of polarization ($P$) of the starlight from the background stars projected towards the outer diffuse envelopes of the clouds, Figure \ref{Fig:Av_map} for the amount of extinction ($A_V$) and Figure \ref{Fig:Pv_Av_map} for the polarization efficiency defined as the amount of polarization produced per unit extinction ($P/A_V$). The contours for IC1396A, BRC37, BRC38, BRC39, LBN437 and GAL110-13 are the WISE 12$\mu$m contours and for L323, L328 and L331, the contours are the Digitized Sky Survey (DSS) 2 contours at R-band. All these parameters vary at different regions of the clouds as expected. $A_V$ is a reliable tracer for the amount of dust concentration along the line of sight, and its variations give us an idea on the varying dust column density along the lines of sight.
$P/A_V$ variations imply the variations in the polarising efficiency of the grains and the grain alignment efficiency in different dust density regions. In the following section, we study the variations of $P$, $A_V$ and polarization angle, $PA$, as functions of the distance $d$ of the stars.

\subsection{\it{Extinction, polarization degree and polarization angle as a function of distance}}\label{subsection:polarization efficiency}

\begin{figure*}
\centering
\resizebox{13.0cm}{7.0cm}{\includegraphics{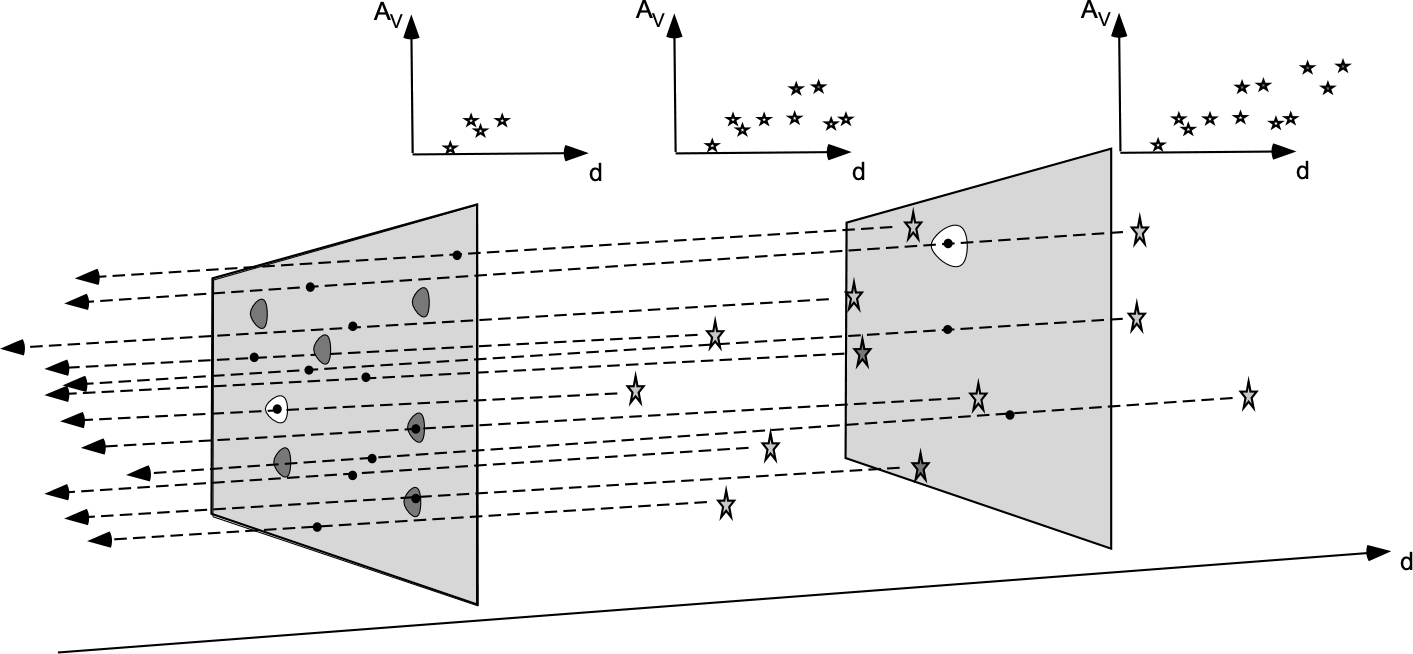}}
\caption{This cartoon is adopted from \cite{2021AJ....161..149S}. The two causes of extinction steps in an extinction vs. distance plot are illustrated in a toy-model/cartoon. For a cloud with a density structure -- especially where the surface area of the high extinction regions is significantly smaller than the average extinction of the cloud/layer -- an extinction jump may be detected at a distance corresponding to where the chance of a line of sight encountering a clump becomes significant.  The lower envelope of the distribution will, however stay fixed. When a second extinction layer is encountered both the upper and lower envelopes of the distribution will change. For extinction layers with low density sub-regions (holes shown as white regions in the screens), a small number of low $A_V$ outliers may be expected. These different behaviors can all be seen in the $A_V$ versus $d$ illustrations shown in the top panel.}\label{Fig:cartoon}
\end{figure*}

Similar to the work of \cite{2021AJ....161..149S} on variation of extinction ($A_V$) with distance ($d$) towards the IC63 and IC59 nebulae, we study the variations of $A_V$ with d for the stars projected towards our target clouds. To isolate the contribution of the dust in the clouds only, and to identify possible additional layers of extinction, we plot the extinction versus distance for the stars.  An illustration of the expected variation of $A_V$ with distance is shown in Figure \ref{Fig:cartoon} (see section \ref{section: discussion on dust layers} for a detailed discussion) While $A_V$ will increase at the distance of the clouds, the polarization degree may not increase in a similar way, because while all grains contribute to the extinction, only non-spherical, paramagnetic grains cause polarization - if aligned. 

A uniform distribution of $PA$ with distance of the stars implies well-ordered magnetic field orientations up to far distances. The variations of $A_V$, $P$ and $PA$ with the distance of the stars are studied for the outer diffuse regions towards each individual cloud as follows.

\begin{figure*}
    \centering
        \includegraphics[scale=0.45]{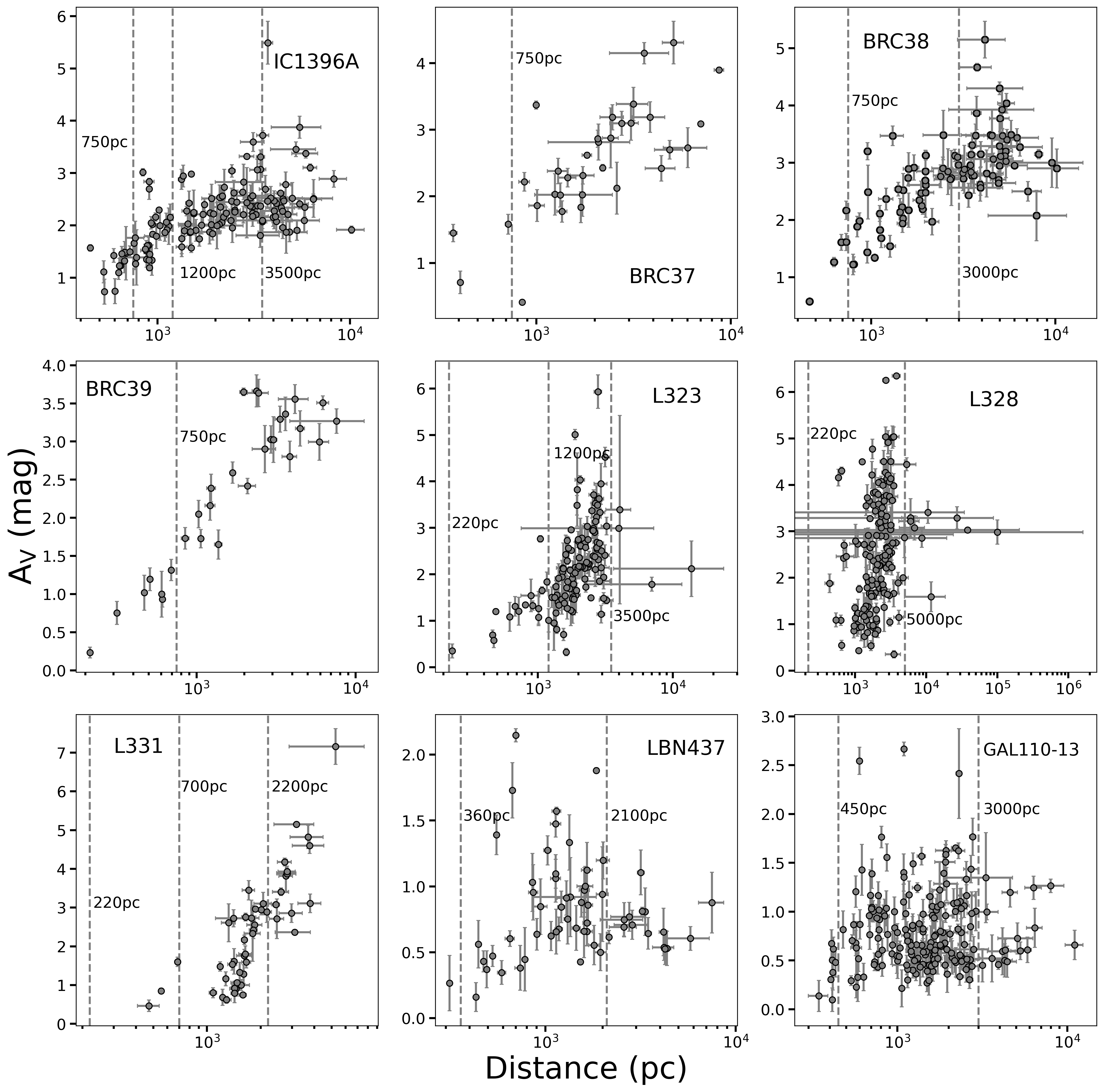}
    \caption{Variations of extinction $A_V$ with distance $d$ of the stars projected towards each of the respective clouds. The first vertical dashed line denotes the estimated distance of the cloud and other dashed line(s) denote(s) the transition distance(s) in the variation pattern of $A_V$ which provide hints on the presence of additional distinct dust layers at those distances.}
    \label{Fig:Av_with_distance}
\end{figure*}

{\bf IC1396A:} The variation of $A_V$ with d for IC1396A shows an overall increase in the extinction with distances (see Figure \ref{Fig:Av_with_distance}). We notice hints of the presence of several layers of dust extinction in the direction of this region. The cloud is at 750 pc and after 750 pc there is a continuous increase in $A_V$ up to a distance of 1200 pc followed by a different increasing variation pattern up to 3500 pc. After 3500 pc, the $A_V$ distribution shows another different pattern, implying the presence of another distinct dust layer. These distances are marked with vertical gray dashed lines.

\begin{figure*}
    \centering
        \includegraphics[scale=0.45]{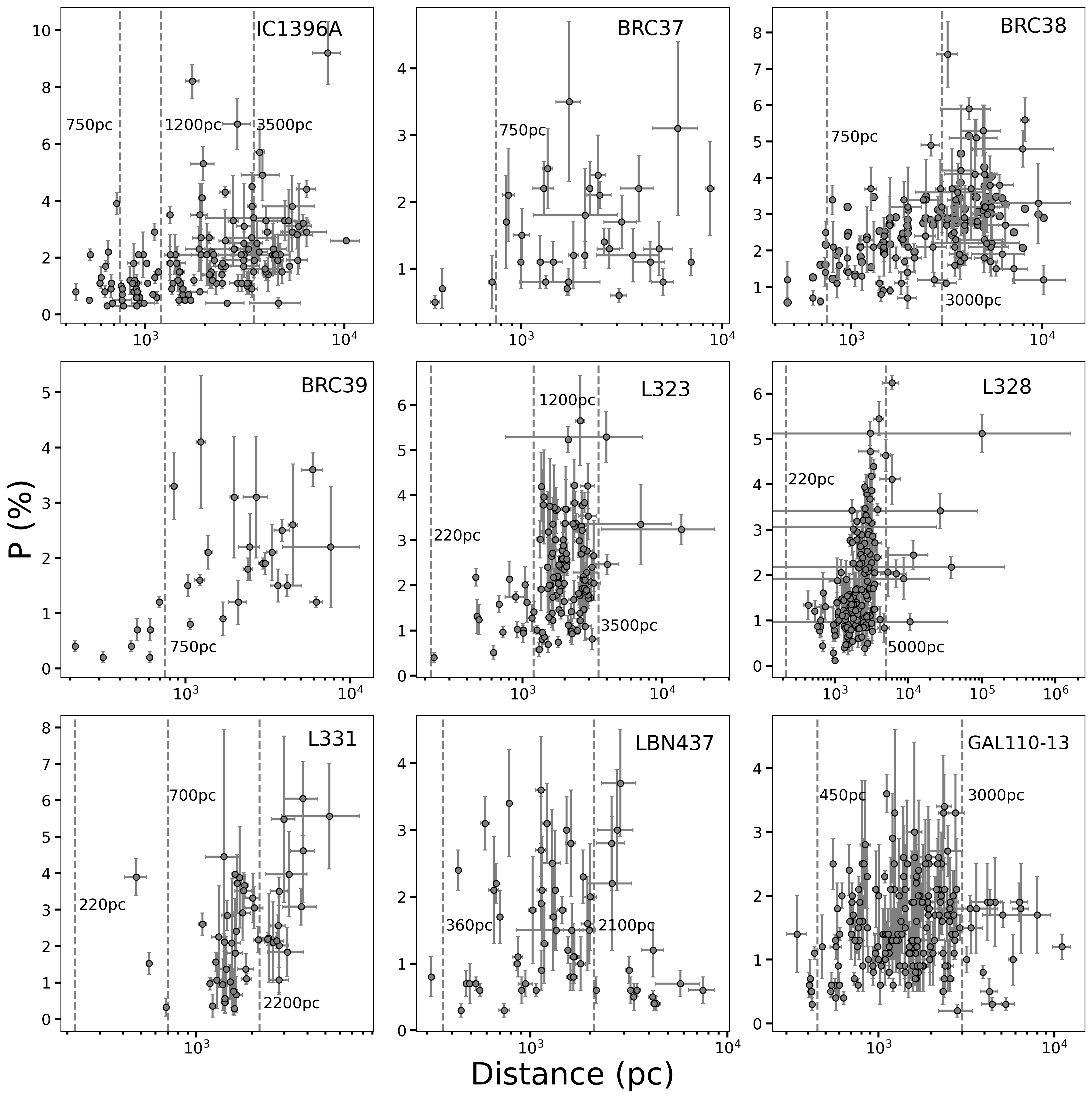}
    \caption{Variations of degree of polarization $P$ with distance $d$ of the projected stars. The vertical dashed lines are marked same as in Figure \ref{Fig:Av_with_distance}.}
    \label{Fig:P_with_distance}
\end{figure*}

As shown in Figure \ref{Fig:P_with_distance}, the polarization degree increases gradually with distance but not at the same rate as $A_V$(d) (see Figure \ref{Fig:Av_with_distance}). 
This can be because of the dependence of $P$ on different parameters like grain properties (such as grain size, shape, composition), magnetic field orientation, angle between the direction of incident radiation and magnetic field and efficiency of grain alignment in different regions etc. 
We mark the same distances as marked in Figure \ref{Fig:Av_with_distance}. As the distance increases, the amount of dust distribution in a column along the line of sight increases. The increase in $P$ with $d$ implies that there are aligned grains at all distances.

\begin{figure*}
    \centering
        \includegraphics[scale=0.46]{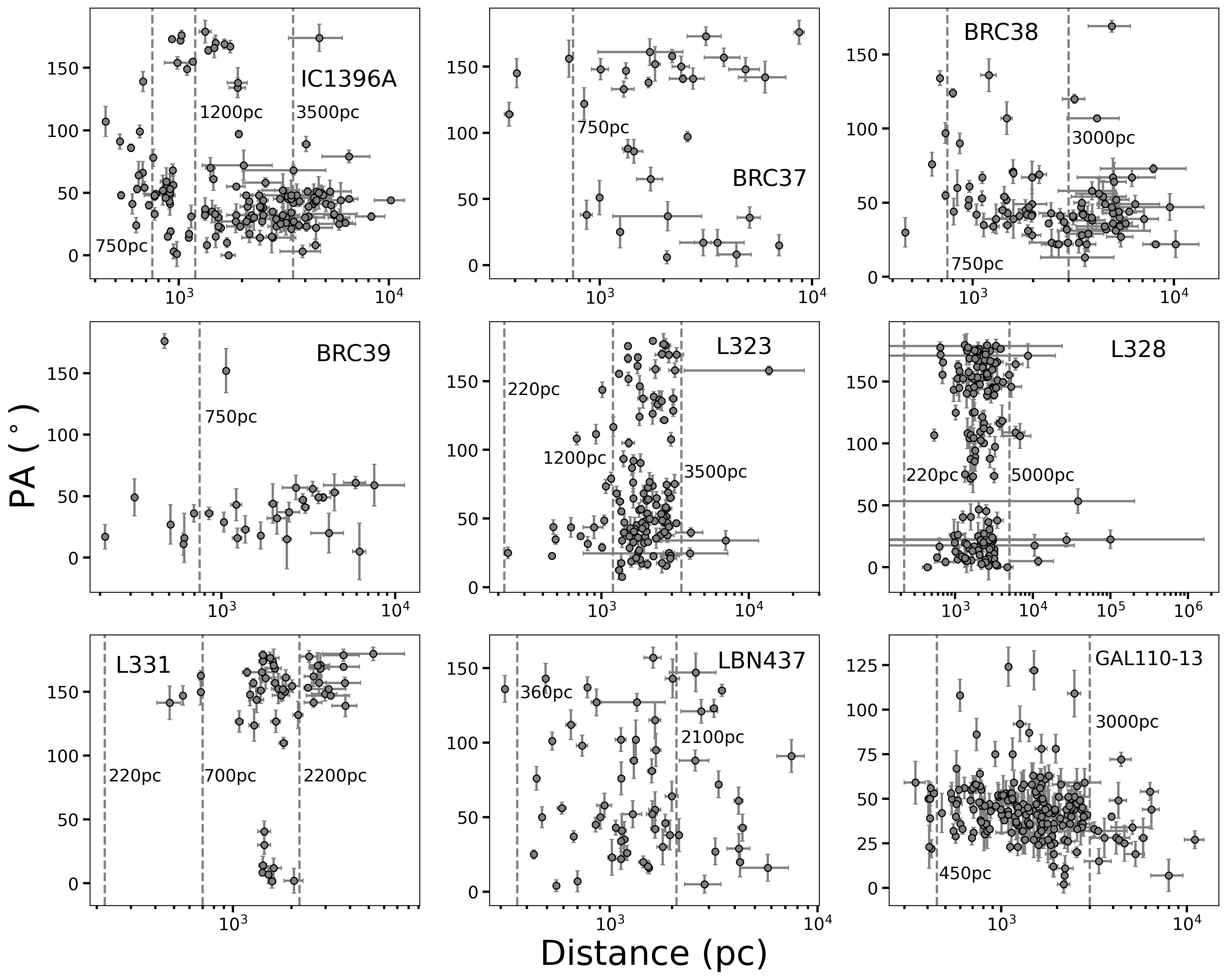}
    \caption{Variations of polarization angle $PA$ with distance $d$ of the projected stars. Same vertical dashed lines are marked as in Figure \ref{Fig:Av_with_distance}.}
    \label{Fig:PA_with_distance}
\end{figure*}

The distribution of polarization position angles with distance (Figure \ref{Fig:PA_with_distance}), show a concentration close to $50^\circ$ with a secondary locus at $\sim 165^\circ$. This is in agreement with the major magnetic field component at $50^\circ$ found by \cite{2018MNRAS.476.4782S}. If we look together the variations of $P$, $A_V$, and $PA$ as a function of distance of the stars, the extinction and the polarization increases with less spread when the distribution of the polarization angles are nearly uniform which may imply that the observed degree of polarization depends on the uniformity of the orientation of magnetic fields.

{\bf BRC 37:} Figure \ref{Fig:Av_with_distance}, shows an overall increase of $A_V$ with distance towards BRC37, suggesting the presence of only one layer of dust in the direction towards BRC37. 

In Figure \ref{Fig:P_with_distance}, the variation of polarization with distance is more random compared to other clouds. However, there is a hint of some increase in $P$ with distance. Figure \ref{Fig:PA_with_distance} shows that some of the polarization angles are more concentrated towards an angle of nearly 150$^\circ$ and some others show a more spread distribution closer to 30$^\circ$ indicating a bimodality in magnetic field orientations. This is seen in the magnetic field orientations shown by \cite{2018MNRAS.476.4782S}.

{\bf BRC 38:} Figure 5 shows two extinction steps at ~750 pc and $\sim$2 kpc.  Beyond 3 kpc additional increases in AV are present, but may indicate clumping in the previous layers. The polarization (Figure \ref{Fig:P_with_distance}) shows a similar behavior as in Figure \ref{Fig:Av_with_distance}. The distribution of polarization position angles (Figure \ref{Fig:PA_with_distance} is concentrated at about 40$^\circ$ with a larger dispersion beyond 3 kpc.

{\bf BRC39:} This cloud shows a more uniformly linear increase in extinction versus distances (Figure \ref{Fig:Av_with_distance}), but if discreet extinction layers are assumed, such can be identified at $\sim$200, 750 and 3000 pc, in agreement with the other Cep OB2 clouds.   The $P$ vs. $d$ (Figure \ref{Fig:P_with_distance}) distribution is consistent with this. The position angles are concentrated around a value of $\approx$ $35^\circ$ with a slight increase beyond $\sim$3 kpc.  \ref{Fig:PA_with_distance}. This equatorial PA corresponds to a Galactic PA close to the Galactic plane.

\begin{figure*}
    \centering
        \includegraphics[scale=0.46]{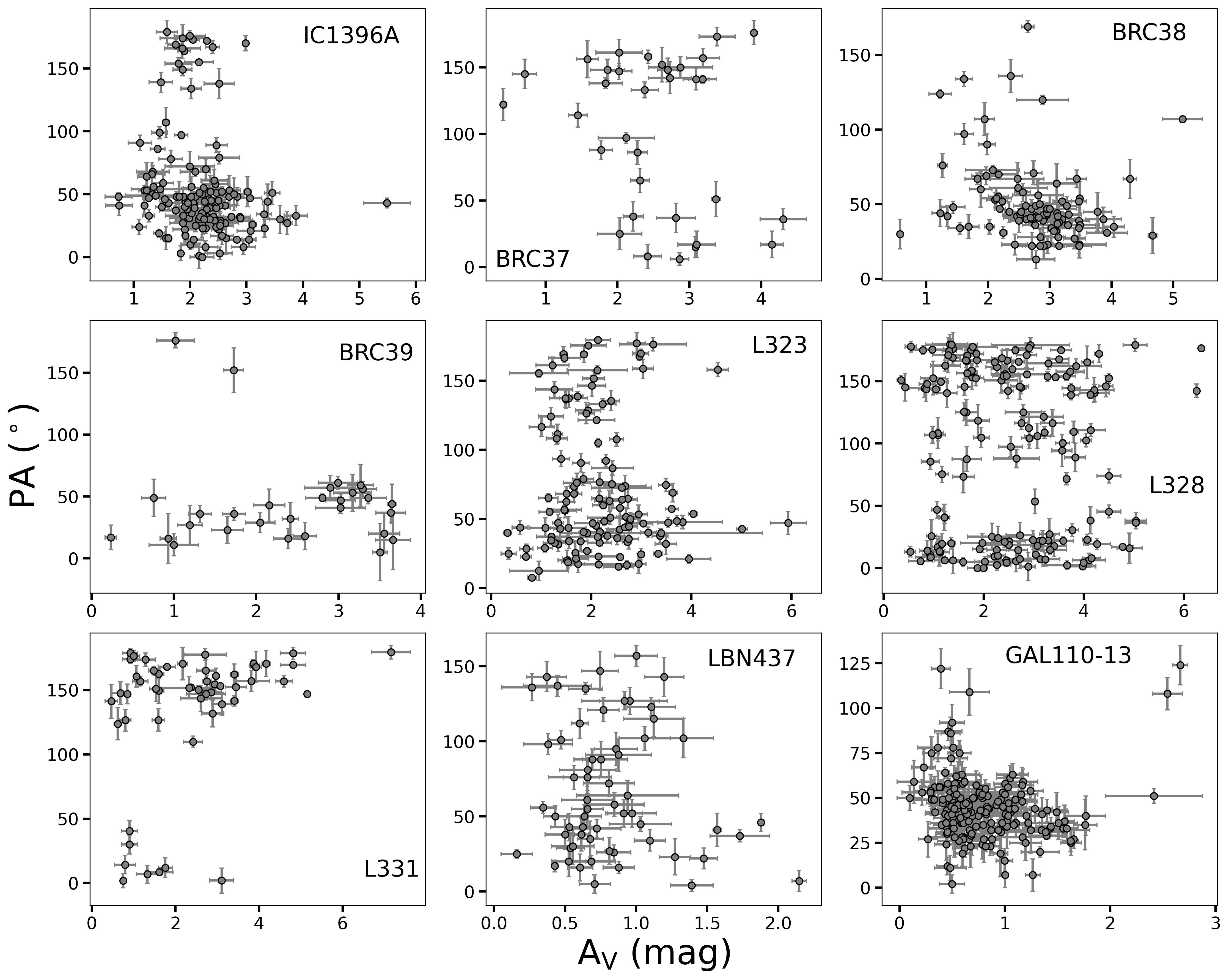}
    \caption{Variations of polarization angle $PA$ with extinction $A_V$ towards each of the clouds.}
    \label{Fig:PA_with_Av}
\end{figure*}

\begin{figure*}
    \centering
        \includegraphics[scale=0.55]{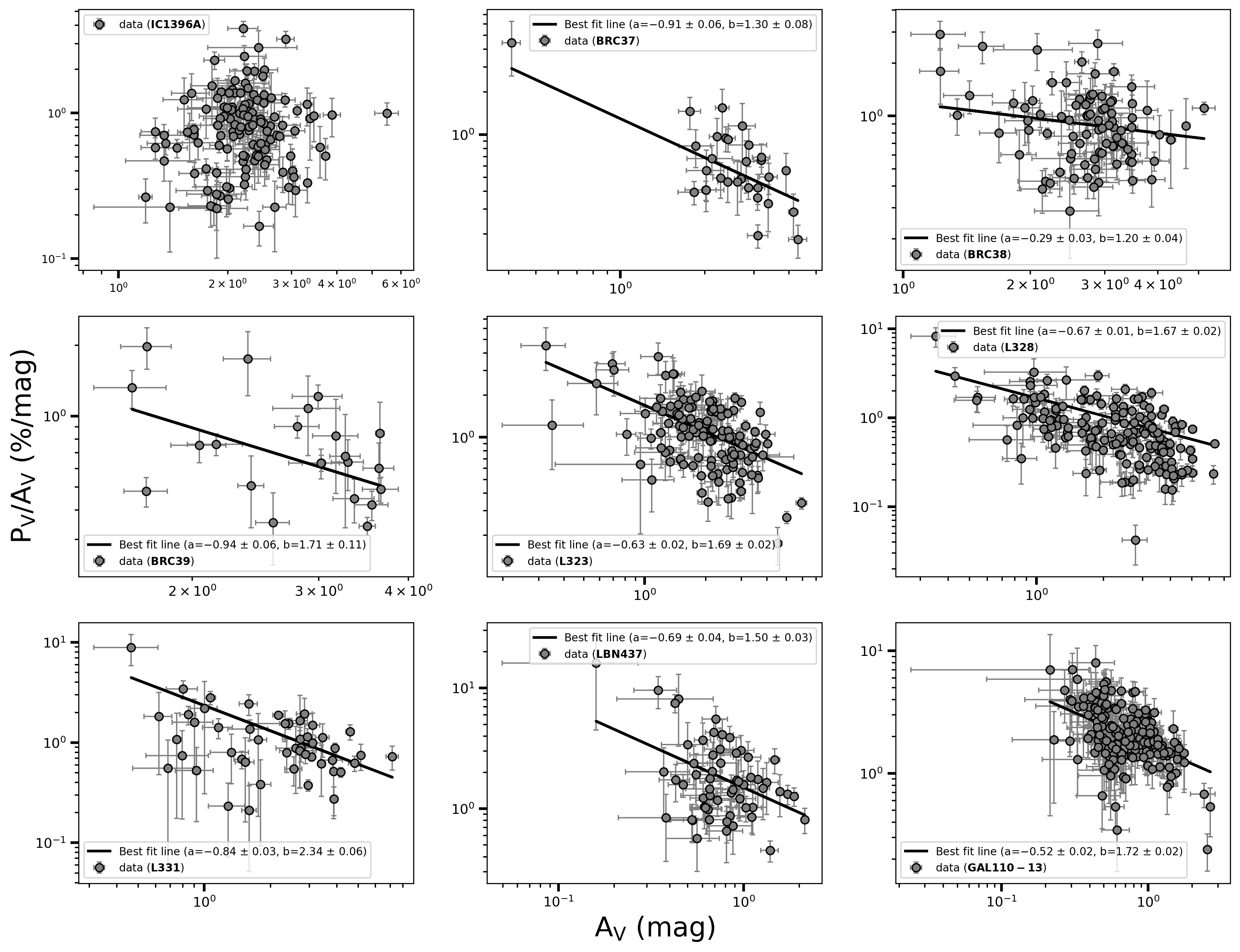}
    \caption{Variations of polarization efficiency ${\frac{P_V}{A_V}}$ with extinction $A_V$ measured towards each of the clouds. The solid black line represents the weighted best fit line of the distribution in log-log scale. The best-fit parameters a and b of the power-law fits of the form $\frac{P_V}{A_V}=b \cdot A_V^a$ for each of the clouds are also given.}
    \label{Fig:Pv_Av_with_Av}
\end{figure*}

\begin{figure*}
\centering
\resizebox{17.0cm}{17.0cm}{\includegraphics{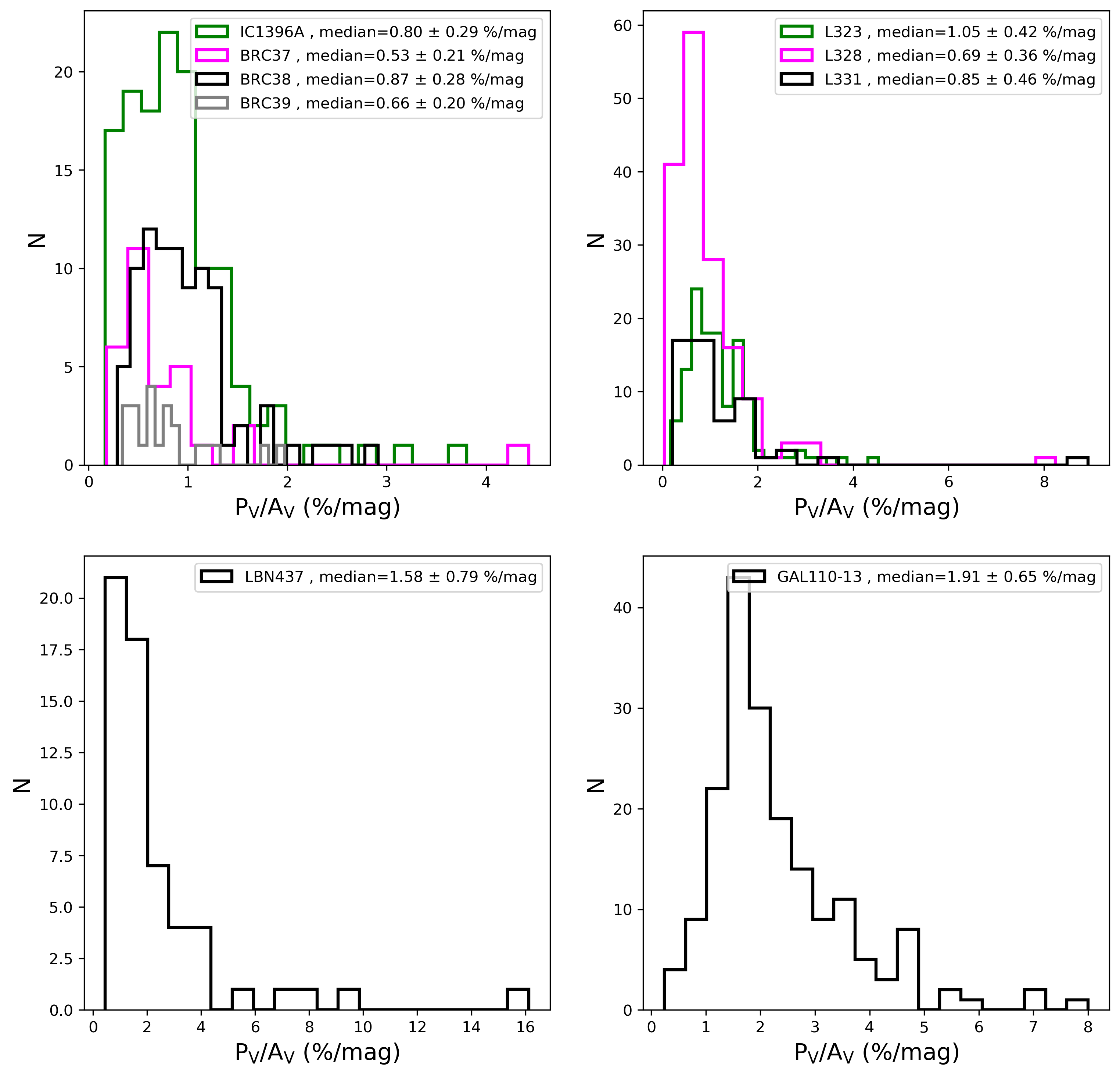}}
\caption{Histograms of polarization efficiency measured towards each of the clouds. The clouds IC1396A, BRC37, BRC38 and BRC39 are grouped together as they are associated with same HII region. Similarly, the clouds L323, L328 and L331 are grouped together as they are found in close proximity to each other.}\label{Fig:Histogram_Pv_Av}
\end{figure*}

{\bf L323:} Figure \ref{Fig:Av_with_distance} shows a shallow increase in $A_V$ with d up to 1200 pc and then changes the variation pattern. After 1200 pc, we see a jump in extinction up to around 3500 pc and then decreases and becomes nearly constant to nearly 2 mag. The variation of polarization with distance also shows a nearly similar trend but with a shallow increase as shown in Figure \ref{Fig:P_with_distance}. This may suggest the presence of more dust column and aligned dust grains with more of the magnetic field in the plane-of-sky, beyond 1200 pc in the direction of L323. We see three distinct groups in the magnetic field orientations or a trimodality nature, with average orientations at around 30, 100 and 150 degree in this layer of dust at the distance between 1200 - 3500 pc as shown in Figure \ref{Fig:PA_with_distance}. This is also seen in magnetic fields morphologies presented by \cite{2023MNRAS.524.1219K}.

{\bf L328:} In Figure \ref{Fig:Av_with_distance} and \ref{Fig:P_with_distance}, the extinction and polarization are found increasing with distance. This is similar behavior as we have seen in L323. The clouds L323, L328, and L331 are the part of same complex. In Figure \ref{Fig:PA_with_distance}, we see a trimodality distribution of position angles which suggests three orientations of magnetic fields at nearly 30$^\circ$, 100$^\circ$ and 150$^\circ$. The magnetic field maps seen in \cite{2015A&A...573A..34S} and \cite{2023MNRAS.524.1219K} also show this behaviour in magnetic field orientations.

{\bf L331:} In the extinction $A_V$ versus distance d plot for L331 as shown in Figure \ref{Fig:Av_with_distance}, we see an increase in the extinction in between 220 pc-700 pc. After 700 pc, the pattern in the extinction changes possibly suggesting the presence of a distinct dust layer. Beyond 2200pc, there is again another increasing pattern in the extinction implying the presence of another distinct dust layer after 2200 pc.

The polarization also increases with distance of stars as shown in Figure \ref{Fig:P_with_distance} nearly similar way but not in the same proportion. $P$ increases from 700-2200 pc and then decreases. After 2200 pc, $P$ again increases in a similar pattern of $A_V-d$ plot shown in Figure \ref{Fig:Av_with_distance}. This may suggest the presence of increasing number of aligned dust grains at all distances. From the polarization angle versus distance plot as shown in Figure \ref{Fig:PA_with_distance}, it is found that most of the polarization angles are concentrated to around 150$^\circ$ which shows that the overall orientations of magnetic fields towards this cloud \citep{2023MNRAS.524.1219K} are almost uniform.

{\bf LBN437:} This is a cometary globule at a distance of around 360 pc. Magnetic fields in this cloud is studied in detail by \cite{2013MNRAS.432.1502S}. The extinction increases with distance of stars up to around 2100 pc and then decreases and becomes nearly constant at around 0.8 mag. The polarization degree also increases but with large spread up to around 3000 pc and then decreases (see Figures \ref{Fig:Av_with_distance} and \ref{Fig:P_with_distance}. The position angle variation with distances (Figure \ref{Fig:PA_with_distance}) is random and it is hard to predict any one orientation of magnetic fields towards this cloud. This chaotic nature of magnetic field geometry is also shown by \cite{2013MNRAS.432.1502S}.

{\bf GAL110-13:} This is another cometary globule at a distance of around 450 pc. The magnetic field properties of this region were studied by \cite{2016A&A...588A..45N}. The $A_V$ and $P$ show similar variation i.e. a slight increase in their values as the distance of the stars increases (see Figures \ref{Fig:Av_with_distance} and \ref{Fig:P_with_distance}).This may suggest the presence of one dust layer behind the cloud. The variation of polarization angle with distance mostly lies at around $\approx$ 40$^\circ$ which may suggest an intact magnetic field orientation up to larger distances. This uniformity in magnetic field orientation can also be seen in \cite{2016A&A...588A..45N}. For the regions towards each of the above different clouds, the variation patterns of $A_V$ with $d$ can be explained by the diagram as shown in Figure \ref{Fig:cartoon}.

\subsection{\it{Polarization angle as a function of extinction}}
The polarization angles here provide the orientations of the plane-of-sky projected magnetic fields. We investigate how $PA$ varies with extinction in the outer diffuse regions towards the different clouds to study the grain alignment efficiency because non-uniform distributions of $PA$ imply magnetic field fluctuations which can decrease the polarization. The variation of $PA$ with $A_V$ is shown in Figure \ref{Fig:PA_with_Av}. It follows nearly similar trend as the variation of $PA$ with $d$ (see Figure \ref{Fig:PA_with_distance}). In IC1396A, BRC37, L323, L328 and L331, we find the presence of more than one magnetic field components.

The polarization angle dispersion (PAD) gives information on the local non-uniformity in the magnetic field morphology \citep{2020A&A...641A..12P}. If PAD is large, it implies that the fluctuations in the magnetic field orientations are large which can reduce polarization.  Small value implies that the magnetic fields are oriented almost uniformly which can increase polarization. Even if the magnetic field orientations are uniform, it does not mean that the polarization will always increase. The polarization depends also on the grain alignment efficiency in the region. If the grains rotate suprathermally and the PAD is small, then the polarization efficiency of the grains will increase as expected by RAT theory (e.g, \citealt{1992ApJ...389..602J}; \citealt{2008MNRAS.388..117H}; \citealt{2016ApJ...831..159H}; \citealt{2020A&A...641A..12P}). Turbulence in the molecular clouds are the main cause of magnetic field fluctuations (\citealt{1989ApJ...346..728J}; \citealt{1992ApJ...389..602J}; \citealt{2008ApJ...679..537F}). In the $PA-d$ and $PA-A_V$ plots shown in Figures \ref{Fig:PA_with_distance} and \ref{Fig:PA_with_Av} respectively, the concentration of $PA$ to some angle or presence of one component ordered magnetic field implies small PAD and for such regions we expect small contribution of magnetic field fluctuations in decreasing the polarization.

\subsection{\it{Polarization efficiency and its variation as a function of extinction}}
The non-spherical aligned grains in the outer parts of the clouds cause polarization of the background starlight. To achieve higher polarization degree, the degree of grain alignment should be high. Along a particular line of sight of a column of dust, all the dust grains there are involved in the extinction. However, to produce polarization only specific grains having properties of non-spherical in shape, silicate-type in composition and sizes greater than 0.05 $\mu$m are responsible. The polarization efficiency $\frac{P_V}{A_V}$ of grains is defined as the degree of polarization produced per unit extinction. It depends on the magnetic field orientations and the grain alignment efficiency along the line of sight. By assuming that the inherent dust properties like roundness, shape, size distribution, composition, etc. at different lines of sights towards a particular cloud are similar, one can use $P$ at $A_V$=1 to study the polarization efficiency of grains along different lines of sights and also its variation in different density regions along different lines of sights. $A_V$ acts as the reliable tracer for the dust concentration and by taking $A_V =1$, we can study how the polarization per unit extinction or the polarization efficiency varies as $A_V$ increases or as we go to denser regions. A decreasing of polarization efficiency with increasing extinction can be caused by the decrease in the alignment efficiency of the grains in the denser regions due to decrease in radiative torques or non-uniform magnetic field orientations along the line of sights or both. RAT theory expects a decreasing trend of grain alignment efficiency as dust extinction increases for uniform orientations of magnetic fields.

In the presence of turbulence, the polarization efficiency as a function of extinction, $A_V$ \citep{1992ApJ...389..602J} is expected to follow a power-law fit of the form 

\begin{equation} 
%\hspace{10em}
{
\frac{P_V}{A_V}=b\cdot{A_V}^{a},
}
%\hspace{10em}
\label{Equation:Pv_Av}
\end{equation}
where the exponent $a$ depends on the turbulence of the medium and grain alignment variation along the line-of-sight (\citealt{2014A&A...569L...1A} ; \citealt{2015AJ....149...31J}) and the parameter $b$ which is equivalent to $\frac{P_V}{A_V}$ at $A_V=1$ depends on the grain alignment efficiency and the orientation of the magnetic field. The value of the exponent $a$ is expected to be $\approx$ $-0.5$ for a fully turbulent medium, and constant grain alignment efficiency indicating a random walk through a large number of turbulent cells with differently oriented magnetic fields \citep{1992ApJ...389..602J}. A steeper dependence on $A_V$ indicates additional mechanisms, likely [partial] loss of grain alignment.

\cite{2019ApJ...873...87M} found the polarization efficiency to be dependent linearly on the intensity or the strength of the external radiation field. A decrease in the radiation strength is expected as the extinction increases and hence a decrease in the polarization efficiency with the increase in extinction.

In plotting the variation of polarization efficiency with extinction as shown in Figure \ref{Fig:Pv_Av_with_Av}, we use the data corresponding to the distance of the projected stars greater than 750 pc for IC1396A, BRC37, BRC38 and BRC39, greater than 220 pc for L323, L328 and L331, greater than 360 pc for LBN437 and greater than 450 pc for GAL110-13 to study only the clouds and the regions behind them (the mentioned distances are the distances of the respective clouds). We plot the weighted best-fit power law in log-log scale describable by Equation \ref{Equation:Pv_Av} and it yields $a=-0.91 \pm 0.06$ and $b=1.30 \pm 0.08$ for BRC37, $a=-0.29 \pm 0.03$ and $b=1.20 \pm 0.04$ for BRC38, $a=-0.94 \pm 0.06$ and $b=1.71 \pm 0.11$ for BRC39, $a=-0.63 \pm 0.02$ and $b=1.69 \pm 0.02$ for L323, $a=-0.67 \pm 0.01$ and $b=1.67 \pm 0.02$ for L328, $a=-0.84 \pm 0.03$ and $b=2.34 \pm 0.06$ for L331, $a=-0.69 \pm 0.04$ and $b=1.50 \pm 0.03$, $a=-0.52 \pm 0.02$ and $b=1.72 \pm 0.02$ for GAL110-13. In each of the clouds except IC1396A, we find a decreasing trend of polarization efficiency with extinction. For IC1396A, the data shows large spread and there is no significant correlation between polarization efficiency and extinction.

The histograms of polarization efficiencies of the grains towards each of the clouds are shown in Figure \ref{Fig:Histogram_Pv_Av}. The clouds of the same complex are grouped together. The upper left panel shows the histograms of $P_V/A_V$ for the regions towards the clouds IC1396A, BRC37, BRC38 and BRC39 with the median values estimated to be $0.80 \pm 0.29$, $0.53 \pm 0.21$, $0.87 \pm 0.28$ and $0.66 \pm 0.20$ \%/mag, respectively. The uncertainties in the median values are estimated from the quartiles. The upper right panel represents for the regions towards L323, L328 and L331 with the median values of $1.05 \pm 0.42$, $0.69 \pm 0.36$ and $0.85 \pm 0.46$ \%/mag, respectively. The lower left and right panels represent for the regions towards LBN437 and GAL110-13 with the median values of $1.58 \pm 0.79$ and $1.91 \pm 0.65$ \%/mag, respectively. For the regions towards each of the clouds, the median polarization efficiencies are overall less than 2 \%/mag with LBN437 and GAL110-13 showing higher polarization efficiencies than the other clouds.

\subsection{\it{Dust Temperature}}\label{subsection:Dust Temperature}
\begin{figure*}
    \centering
    \begin{tabular}{ccc}
        \includegraphics[scale=0.45]{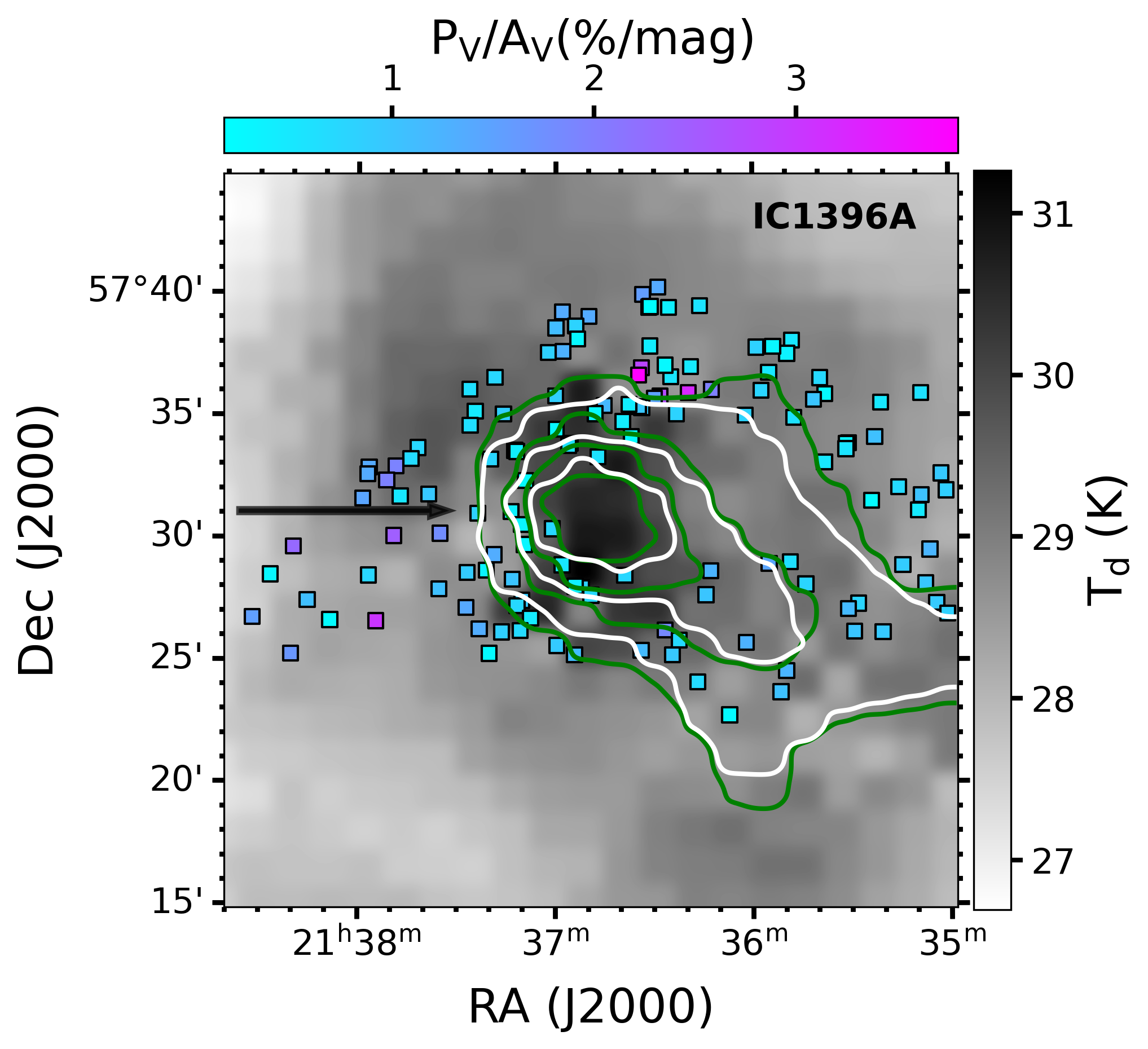} &
        \hspace{-15pt} % Adjust this value to reduce or increase the gap
        \includegraphics[scale=0.45]{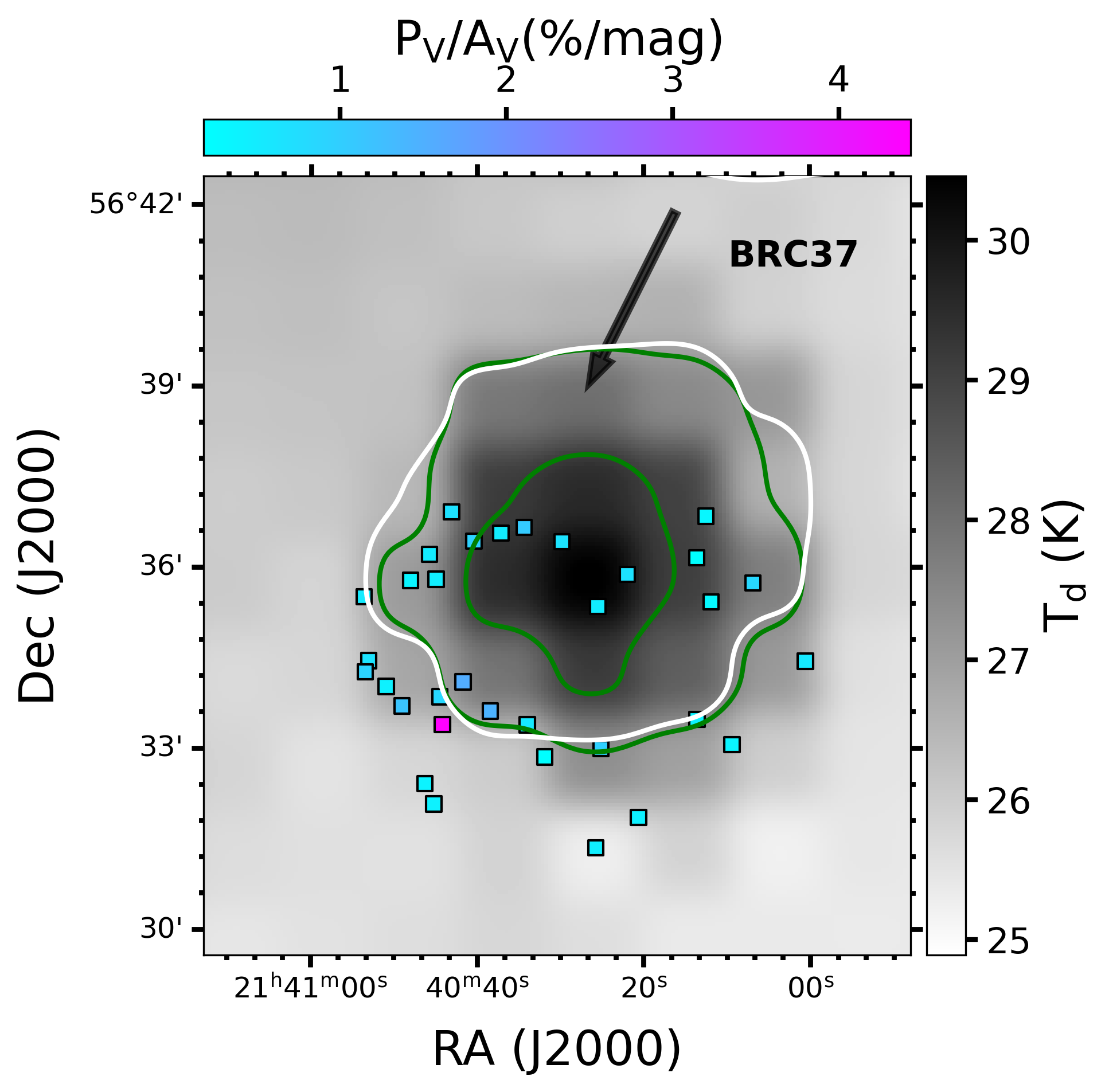}\\
        \hspace{-15pt} % Adjust this value to reduce or increase the gap
        \includegraphics[scale=0.45]{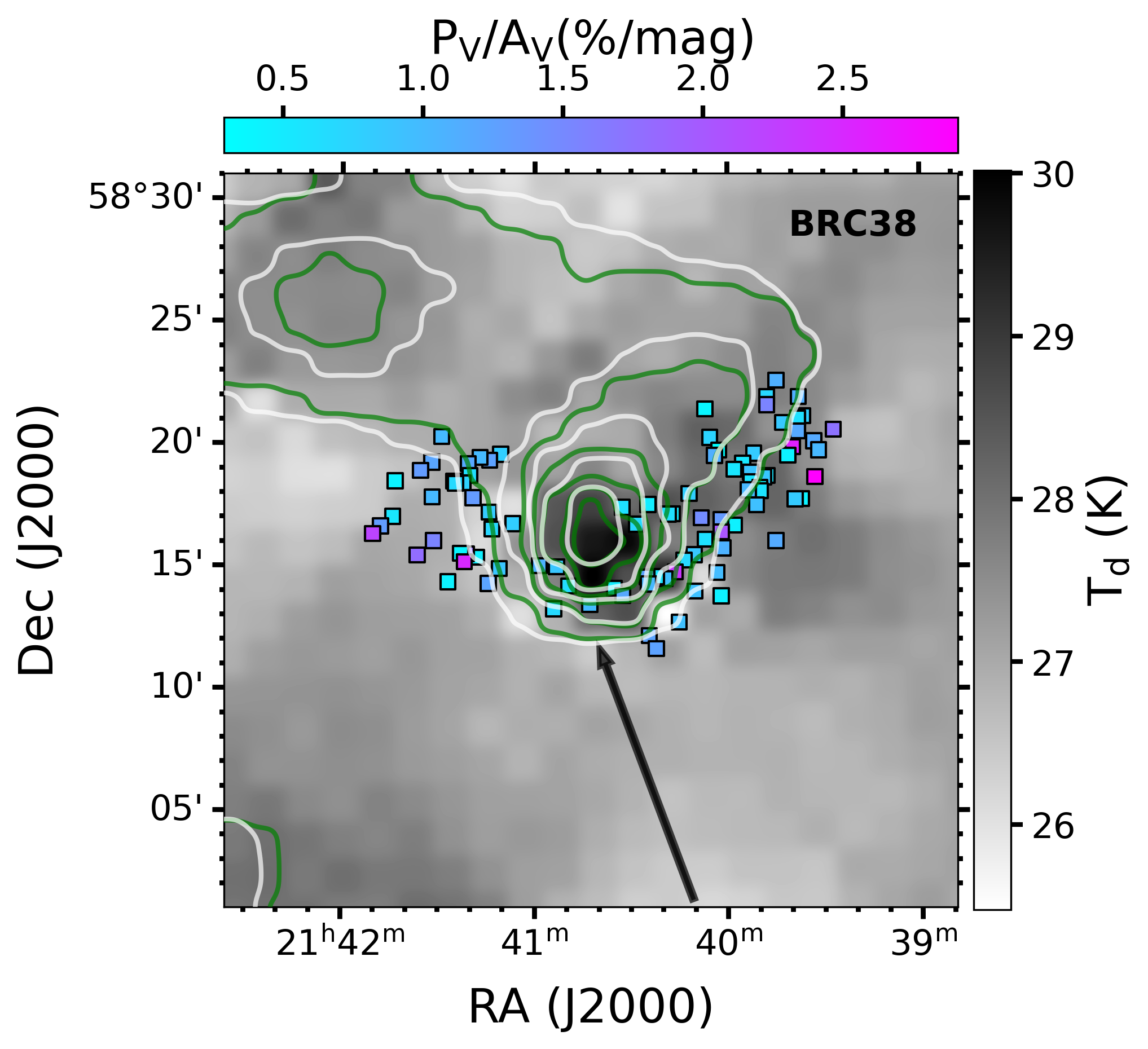} &
        \hspace{-15pt} % Adjust this value to reduce or increase the gap
        \includegraphics[scale=0.45]{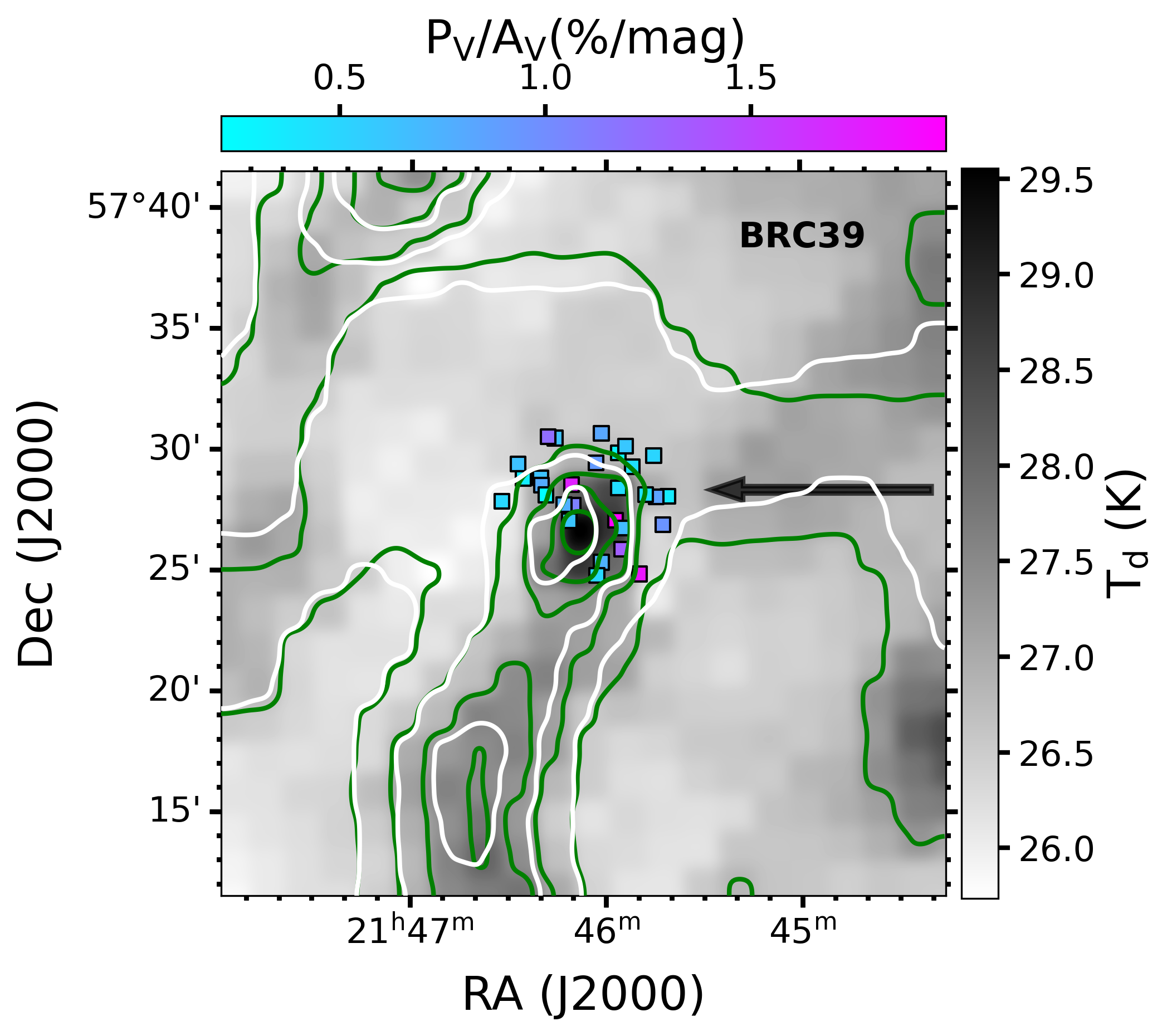}\\
        \hspace{-15pt} % Adjust this value to reduce or increase the gap
        \includegraphics[scale=0.48]{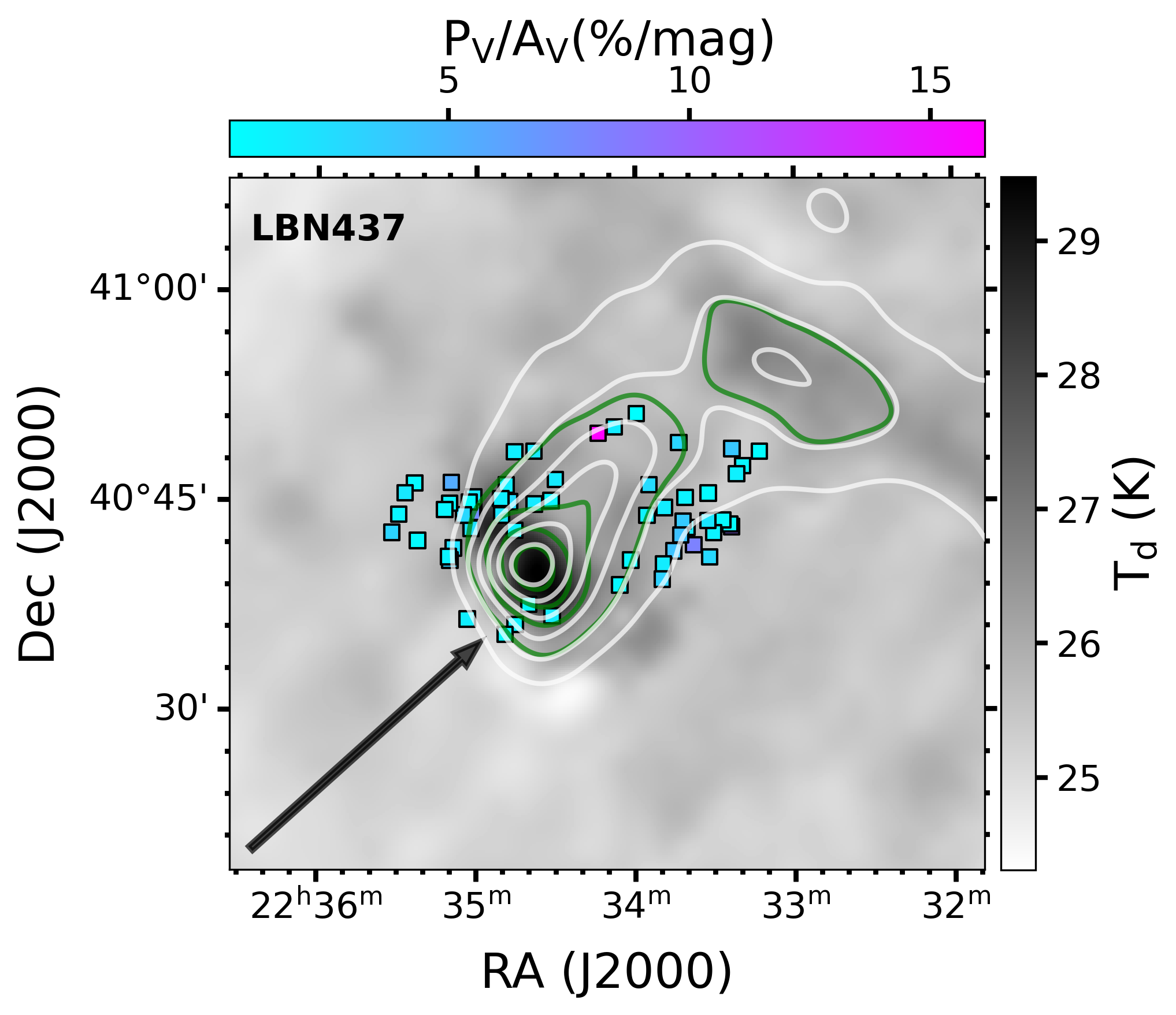} &
        \hspace{-15pt} % Adjust this value to reduce or increase the gap
        \includegraphics[scale=0.45]{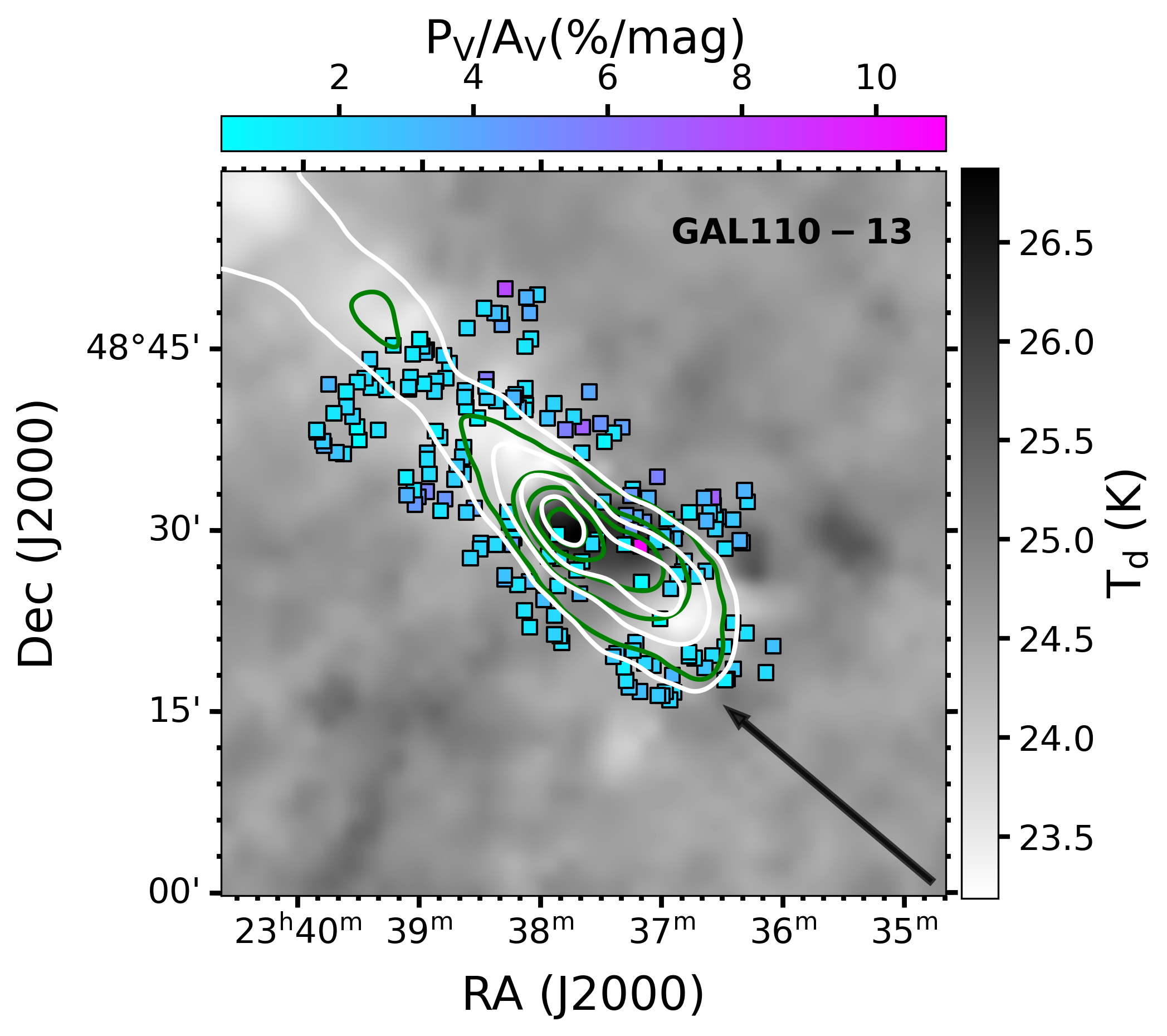}
    \end{tabular}
    \caption{Dust temperature maps for the clouds IC1396A, BRC37, BRC38, BRC39, LBN437 and GAL110-13 made using IRIS 60 $\mu$m and 100 $\mu$m data. The green and the white contours represent the IRIS 60 $\mu$m and 100 $\mu$m contours, respectively for each of these clouds. The polarization efficiency values are also overplotted on these dust temperature maps. The arrows denote the radiation directions from the ionizing stars of the clouds.}
    \label{Fig:Td_map}
\end{figure*}

From RAT theory, a positive correlation between the degree of polarization and radiation strength or equivalently dust temperature is expected, because both grain alignment and heating is driven by the radiation field strength (\citealt{2007MNRAS.378..910L} ; \citealt{2015ARA&A..53..501A}). To see whether this correlation is found, we make dust temperature maps of IC1396A, BRC37, BRC38, BRC39, LBN437 and GAL110-13 using Improved Reprocessing of the IRAS Survey (IRIS) data where IRAS denotes the Infrared Astronomical Satellite. The dust temperature maps are shown in Figure \ref{Fig:Td_map} with the green contours representing IRIS 60 $\mu$m emission and white contours IRIS 100 $\mu$m emission. The dust temperature is derived assuming that the dust in each pixel is isothermal and that the ratio of 60 to 100 $\mu$m dust emissions follows a blackbody function of temperature $T_\mathrm{d}$, modified by a power-law emissivity spectral index $\beta$ \cite{2005ApJ...634..442S}:

\begin{equation}
\hspace{7em}
{
{T_\mathrm{d}=\frac{-96}{ln\left\{R \times 0.6^{(3+\beta)}\right\}}},
}
\hspace{7em}
\end{equation} 
where $R$ is the ratio of flux densities at 60$\mu$m and 100$\mu$m, $\beta$ is the emissivity spectral index of the dust and $T_\mathrm{d}$ is the dust temperature at each pixel of the map. The value of $\beta$ depends on the properties of the dust grains like their composition, size, etc. Its value is taken to be zero for a pure blackbody. The amorphous layer-lattice matter has the value of $\beta$ nearly 1 and the metals and crystalline dielectrics have the value of $\beta$ nearly 2. We consider silicate grains and take $\beta$ value equal to 1.8 \citep{2005ApJ...634..442S} as the silicate materials have dielectric nature to derive the dust temperature for all the clouds.

After deriving the temperature maps, we match the position of the stars having polarization and extinction measurements with the pixel coordinates in the temperature maps. Then we calculate the mean value in $T_\mathrm{d}$ for all the pixel values within a diameter of $2'$ around each of these pixel coordinates of the stars as the resolutions of IRIS 60 and 100 $\mu$m are $2'$.

\begin{figure*}
    \centering
        \includegraphics[scale=0.68]{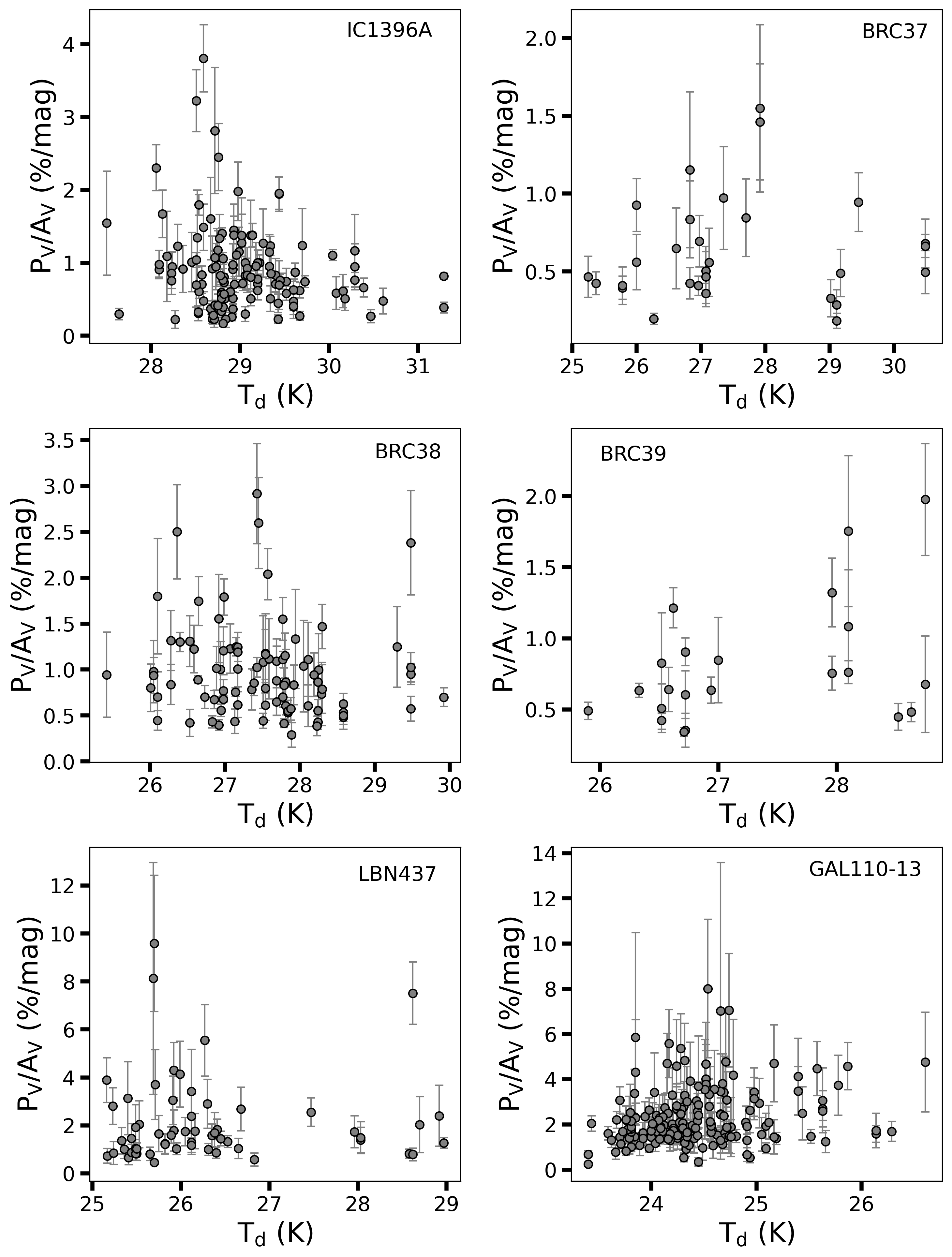}
    \caption{Variations of polarization efficiencies $\frac{P_V}{A_V}$ with increasing dust temperature $T_d$ towards the clouds IC1396A, BRC37, BRC38, BRC39, LBN437 and GAL110-13.}
    \label{Fig:Pv_Av_with_Td}
\end{figure*}

We investigate the variation of polarization efficiency with the dust temperature. This variation is shown in Figure \ref{Fig:Pv_Av_with_Td}. In IC1396A, BRC37, BRC38, BRC39 and GAL110-13, the data points show large spread overall. However, we see some increase in $P$. In LBN437, the polarization efficiency is almost constant with increasing $T_\mathrm{d}$. For the regions of very ordered magnetic field orientations along the line-of-sight, we expect increasing of the polarization efficiency with the dust temperature or the radiation strength in the context of RAT theory (a parallel work on testing this in all these clouds, is under preparation). If the magnetic field orientations are not ordered, it can reduce the polarization efficiency.

\begin{table*}
	\centering
	\caption{This table gives the value of the best fit parameters $a$ and $b$ in the logarithmic scale plot of $\frac{P_V}{A_V}$ with $A_V$ towards the different clouds with the power law fits given by $\frac{P_V}{A_V}=b \cdot A_V^a$ with the first section for the clouds under this study and the second section is for the clouds from previous studies.}
	\label{table:slope_intercept}
	\begin{tabular}{lcccr} % four columns, alignment for each
            \hline
		Cloud & a & b & Median $\mathrm{P_V/A_V}$ (\%/mag)\\ 
		\hline
		IC1396A & - & - & 0.80 $\pm$ 0.29\\
            BRC37 & $-0.91 \pm 0.06$ & 1.30 $\pm$ 0.08 & 0.53 $\pm$ 0.21\\
		BRC38 & $-0.29 \pm 0.03$ & 1.20 $\pm$ 0.04 & 0.87 $\pm$ 0.28\\
            BRC39 & $-0.94 \pm 0.06$ & 1.71 $\pm$ 0.11 & 0.66 $\pm$ 0.20\\
            L323 & $-0.63 \pm 0.02$ & 1.69 $\pm$ 0.02 & 1.05 $\pm$ 0.42\\
            L328 & $-0.67 \pm 0.01$ & 1.67 $\pm$ 0.02 & 0.69 $\pm$ 0.36\\
		L331 & $-0.84 \pm 0.03$ & 2.34 $\pm$ 0.06 & 0.85 $\pm$ 0.46\\
            LBN437 & $-0.69 \pm 0.04$ & 1.50 $\pm$ 0.03 & 1.58 $\pm$ 0.79\\
            GAL110-13 & $-0.52 \pm 0.02$ & 1.72 $\pm$ 0.02 & 1.91 $\pm$ 0.65\\
            \hline
            \hline
            IC63\textsuperscript{$\ast$} & $-0.55 \pm 0.09$ & 2.35 $\pm$ 0.04 \\
		  IC59\textsuperscript{$\ast$} & $-0.41 \pm 0.08$ & 1.88 $\pm$ 0.04 \\
            LDN204\textsuperscript{$\ast$} & $-1.24 \pm 0.07$ & 3.72 $\pm$ 0.04 \\
            \hline
            \hline
            \textsuperscript{$\ast$} denotes previously studied clouds taken from \cite{2021AJ....161..149S}.
	\end{tabular}
\end{table*}

\section{Discussions}\label{section:Discussions}

\subsection{\it{Evidence of presence of dust layers towards the clouds: $A_V$ versus d}}\label{section: discussion on dust layers}
The study of the variation patterns of extinction with increasing distances of the background stars projected in the direction towards the clouds helps us to get information on the presence of additional distinct dust layers at certain ranges of distance. If we consider a uniform space distributions of background stars, their surface density on the sky will be the product of their space density and the distance. So, their surface density will rise linearly with distance. With increasing line-of-sight distance, there will be decrease in the on-the-sky distance between observed stars.

For an extended low-extinction layer having small dense sub-regions of higher extinction values inside a localised region, it is expected that there will be sudden increase in the upper envelope of the extinction corresponding to a single scale size of the dense sub-regions or there will be gradual increase in the upper envelope of the extinction for a distributions of high-extinction clumps. Considering the large scale extended low-extinction part the lower envelope is expected to be nearly constant. If there is a new distinct extinction layer then the lower envelope of the extinction is expected to rise.

We can also consider a different interstellar density distributions where there is more extended distributions of high-extinction dense regions with smaller lower-extinction less dense regions. In this situation also, both the upper and lower envelopes of the extinction will rise at the characteristic distance with less number of lower extinction points remaining (details are given in \cite{2021AJ....161..149S}). This discussion is explained in Figure \ref{Fig:cartoon}.

The significant change in the $A_V$ distribution in some of the clouds like IC1396A, BRC38, L323, L331 (see Figure \ref{Fig:Av_with_distance}) at certain distances suggests the presence of additional dust layers at those distances. The very significant increase in $A_V$ in L323 after 1200pc implies a new more dense dust layer up to 3000 pc.

\subsection{\it{Polarizing properties of the dust grains towards the clouds}}
The degree of polarization of starlight by dust grains in molecular clouds depends on the properties of dust grains like their sizes, shapes, compositions etc., their alignment efficiency along the line-of-sight and the line-of-sight fluctuations in the magnetic field orientations. As the distance of the background star increases, the starlight will traverse through more dust in a column along the line-of-sight, thereby causing more extinction but polarization degree due to the same grains may not follow the similar variation with distance due to its dependence on the above mentioned parameters. If the fraction of aligned grains along the dust column in a more uniform magnetic field increases, then the overall polarization will increase. In our study, the increase in $P$ with $d$ of stars up to far distances behind the clouds (see Figure \ref{Fig:P_with_distance}) may imply the presence of aligned dust grains at all distances and the dust layers present at the particular range of distances effectively polarize the starlight, thereby acting as polarizing layers.

\subsection{\it{Polarization efficiencies of the grains towards the BRCs and CGs}}
We compare the polarization efficiencies of the grains in the outer diffuse envelopes of the BRCs and CGs by grouping into clouds of same HII region illuminated by same O/B-type stars (see Figure \ref{Fig:Histogram_Pv_Av}). The BRCs IC1396A, BRC37, BRC38, and BRC39 are located in different directions of same HII region IC1396 and the median polarization efficiencies of grains towards these clouds are found similar. Similarly, L323, L328, and L331, which are part of same complex shows nearly similar median polarization efficiencies of the grains towards their outer regions. For the isolated CGs LBN437 and GAL110-13, they are illuminated by the same star 10 Lac and the median polarization efficiencies are nearly similar with values nearly 2 \%/mag. The $A_V$ values in our study ranges up to $\approx$ 4 mag. A couple of clouds have $A_V$ values up to 6 mag and beyond. The similarity in polarization efficiency in clouds of same group may be explained by their similar environments and illumination by the same ionizing source.

\subsection{\it{Implications for grain alignment by RATs}}
\subsubsection{Decrease of polarization efficiency with extinction}
The most acceptable theory of grain alignment is the alignment induced by RAdiative Torques (RATs) known as RAT theory. Many observational results in different studies in different regions have been well explained by this theory. RAT theory works better for larger grains of size $> 0.05$ $\mu$m (for a review see \citealt{2015ARA&A..53..501A}).
The alignment efficiency of grains in molecular clouds is found to decrease with increasing dust extinction or in denser regions which is expected from RAT theory in many studies (e.g \citealt{2019ApJ...873...87M} ; \citealt{2021AJ....161..149S}) when the cloud does not have internal radiation sources and the source of radiation is from external ionizing source and diffuse interstellar radiation field. In our study, the decrease in the polarization efficiency with extinction, as observed in regions of less PAD towards BRC38, BRC39, L331 and GAL110-13 (see Figure \ref{Fig:Pv_Av_with_Av}) can be due to the decrease in RAT alignment efficiency of grains in denser regions. However, magnetic field fluctuations along the lines of sights may have significant contribution for the decrease in polarization efficiency with extinction in the regions towards BRC37, L323 and L328. For the regions towards LBN437 the decrease in polarization efficiency with extinction may be majorly due to magnetic field tangling because of its non-uniform magnetic field orientations.

\subsubsection{Polarization efficiency variation with dust temperature}
From RAT theory, it is expected that the polarization efficiency should increase with the increase in the strength of radiation field or equivalently dust temperature for uniform magnetic field orientations along the line-of-sight (e.g, \citealt{2007MNRAS.378..910L}; \citealt{2015ARA&A..53..501A}; \citealt{2019ApJ...873...87M}; \citealt{2021ApJ...908..218H}). In our study of the variation of polarization efficiency with dust temperature (see Figure \ref{Fig:Pv_Av_with_Td}), the data points show large spread in each of the regions towards the clouds. In LBN437, the polarization efficiency is found to be almost constant with $T_d$. For the clouds IC1396A, BRC38, BRC39 and GAL110-13 which have ordered distribution of magnetic field orientations up to far distances (see Figure \ref{Fig:PA_with_distance}), we get some hints of increase in P with $T_d$, which may imply the alignment of grains by RATs. However, for the region towards the cloud LBN437 which has more randomness in magnetic field orientations, the constant variation of the polarization efficiency can be due to the non-uniform distribution of magnetic field orientations which can reduce the polarization efficiency.

\subsection{\it{Comparison with previous studies}}\label{subsection:Comparision}

In the study of polarization efficiency in IC 59, IC63, LDN 204 \citep{2021AJ....161..149S} and in Local Bubble Wall \citep{2019ApJ...873...87M}, these regions show decrease in the polarization efficiency as the dust extinction increases which is in good agreement with RAT alignment mechanism of dust grains. The decreasing trend of polarization efficiency with the increase in extinction describable by a power law decay has also been found in several studies 
(e.g, \citealt{2014A&A...569L...1A}; \citealt{2015ARA&A..53..501A}; \citealt{2015AJ....149...31J}; \citealt{2022MNRAS.513.4899S}; \citealt{2024AJ....167..242S}). In our study of other environments (BRCs and CGs) IC1396A, BRC37, BRC38, BRC39, L323, L328, L331, LBN437 and GAL110-13, it is also found that the polarization efficiency decreases as the dust extinction increases in all these clouds except IC1396A which shows a nearly flat distribution overall. Our study agrees with the observed trends in the previous studies.

In the polarization efficiency vs. extinction study in the Sh 2-185 (IC 59 and IC 63) reflection nebulae by \cite{2021AJ....161..149S}, the slopes were found to be -0.55 $\pm$ 0.09 and -0.41 $\pm$ 0.08 for IC63 and IC59, respectively. The intercepts were obtained as 2.35 $\pm$ 0.04 and 1.88 $\pm$ 0.04 for IC63 and IC59, respectively. They reported these values in cloud 3 of LDN 204 as -1.24 $\pm$ 0.07 slope and 3.72 $\pm$ 0.04 intercept. Table \ref{table:slope_intercept} summarises and compares the slope and intercepts of previously studied clouds by \cite{2021AJ....161..149S} and those of ones studied in this work.

\section{SUMMARY}\label{section:Summary}
In this work, we use optical polarimetric archival data for the background stars projected in the directions towards nine Bright-Rimmed clouds and Cometary Globules, along with the extinction and distance data from StarHorse 2 catalogue. We analyse the data to study the polarization efficiencies of the dust grains towards these clouds and the grain alignment mechanisms. The main findings of our work are summarized as follows:

\begin{itemize}
    \item We analyse the distributions of extinction, polarization degree, and polarization angle with increasing distance of the stars projected towards each of the  clouds. We find that there are implications of presence of multiple dust layers towards the clouds IC1396A, L323, and L331. An increase in the polarization degree with distance of the stars is observed towards each cloud, which may imply the presence of aligned grains at all distances.
    
    \item We estimate the polarization degree produced per unit extinction, termed as polarization efficiency for the grains in the direction towards the outer diffuse envelopes of each of the clouds. We find nearly similar median polarization efficiencies in the clouds of same complex.
    
    \item We find polarization efficiency decreases as extinction increases in most of the clouds. The variations of the polarization efficiency with dust temperature show a large spread overall in almost all the clouds, however some increase can also be seen. We get some hints on the alignment of grains by radiative torques through the study of the variations of the polarization efficiencies with extinction and dust temperature, towards BRC38, BRC39, L331 and GAL110-13 which have almost uniform magnetic field orientations.
      
\end{itemize}

$Software$: APLpy (\citealt{2012ascl.soft08017R} ; \citealt{aplpy2019}), Astropy (\citealt{2013A&A...558A..33A}; \citealt{2018AJ....156..123A}), SciPy \citep{2020NatMe..17..261V}.

$Facilities$: ARIES Imaging Polarimeter archival data

\section{acknowledgments}
I express my sincere gratitude to Dr. Archana Soam and Dr. B.G-Andersson for their useful comments that significantly helped a lot in the improvement of the manuscript. This research has made use of the optical polarization data from the ARIES Imaging Polarimeter (AIMPOL), and also the SIMBAD database, operated at CDS, Strasbourg, France. I also acknowledge the use of NASA’s SkyView facility (http://skyview.gsfc.nasa. gov) (\href{http://skyview.gsfc.nasa.gov}{Sky View}) located at NASA Goddard Space Flight Center.

\bibliography{ms2026-0164_Polarization_references}{}
\bibliographystyle{aasjournal}

\end{document}